\documentclass[final,5p,times,twocolumn]{elsarticle}

\usepackage{amssymb}
\usepackage{amsmath}
\usepackage{comment}
\usepackage{amsthm}
\usepackage{xurl}
\usepackage[colorlinks,allcolors=blue]{hyperref}
\usepackage[T1]{fontenc}
\usepackage[utf8]{inputenc} 
\usepackage{xcolor}
\usepackage{makecell}
\usepackage{multirow}

\usepackage{lineno}
\journal{Astroparticle Physics}

\begin{document}
\begin{frontmatter}



\title{\textcolor{black}{The impact of plasma instability cooling on intergalactic magnetic field constraints in GeV cascades for optimized instability cooling parameters}}

\author[label1]{Suman Dey\corref{cor1}}
\ead{suman.dey@desy.de}

\author[label1]{Simone Rossoni}
\ead{simone.rossoni@desy.de}

\author[label1]{G\"{u}nter Sigl}
\ead{guenter.sigl@desy.de}

\cortext[cor1]{Corresponding author: Suman Dey}
\affiliation[label1]{organization={II. Institut f\"{u}r Theoretische Physik, Universit\"{a}t Hamburg},
            addressline={Luruper Chaussee 149}, 
            city={Hamburg},
            postcode={22761},
            country={Germany}}

\begin{abstract}
Electromagnetic cascades are initiated by TeV gamma rays propagating through the intergalactic medium (IGM), and they can be used to constrain the weak intergalactic magnetic field (IGMF) in cosmic voids. Primary TeV photons produce electrons and positrons through electromagnetic pair production, which can be deflected out of the line-of-sight to the observer by IGMF. In addition, electron-positron pairs can perturb the IGM, triggering plasma instabilities that can cool down the pairs before they upscatter cosmic background photons to GeV energies via inverse Compton (IC) scattering. In this work, we investigate the influence of plasma instabilities on the cascade spectrum by introducing a parameterized instability model within the publicly available Monte Carlo framework \texttt{CRPropa 3.2} in the presence of IGMF. We first determine the instability parameters that best reproduce the Fermi-LAT observations in the absence of any IGMF. We then use extended-emission observations within the observer’s field of view, including the effects of the IGMF, to constrain the IGMF strength in the presence of the corresponding best-fit instability-cooling parameters, based on the Fermi-LAT spectral observations of the blazar 1ES 0229+200. We find that plasma instabilities with a characteristic length scale of $120~\text{kpc}$ and a spectral index of $\alpha=-0.5$ are consistent with the observed photon spectra. We also find that the fit of the observed data is improved by the presence of an IGMF: we obtain an IGMF lower limit of $B \gtrsim 2.7 \times 10^{-17}~\text{G}$ for an observer field of view $1.0^\circ$.
\end{abstract}


\end{frontmatter}


\section{Introduction}
\label{sec:intro}
Blazars are a class of Active Galactic Nuclei (AGN) characterized by a TeV gamma-ray jet that is aligned closely with our line of sight. As the emitted TeV photons propagate through the intergalactic medium (IGM), they interact with the extragalactic background light (EBL) \cite{Hauser:2001xs}, which is optically thick for photons at TeV energies. The result of the interaction is the production of $e^{+}e^{-}$ pairs via the process $\gamma + \gamma_{\text{EBL}} \rightarrow e^+ + e^-$ (i.e. electromagnetic pair production). These electron-positron pairs are produced with Lorentz factor $\Gamma$ typically ranging from $10^{5}$ to $10^{8}$ \cite{Lee:1996fp}. These pairs can subsequently interact with cosmic microwave background (CMB) photons \cite{gould1967opacity, blumenthal1970bremsstrahlung} via inverse Compton (IC) scattering $e^{\pm} + \gamma_{\text{CMB}} \rightarrow e^{\pm} + \gamma$. As a result, lower-energy secondary photons are produced, leading to the development of an electromagnetic cascade \cite{Aharonian:1993vz,Berezinsky:2016feh}. Higher-order processes, such as double pair production (DPP, $\gamma+\gamma \rightarrow e^+ + e^- + e^+ + e^-$) and triple pair production (TPP, $e^\pm + \gamma \rightarrow e^\pm + e^+ + e^-$), are also possible, but they are generally suppressed below energies of about $100\text{~TeV}$ \cite{Heiter:2017cev}.

Measurements of the photon flux produced by electromagnetic cascades are performed either directly with space-based detectors, such as the Fermi Large Area Telescope (Fermi-LAT) \cite{Fermi-LAT:2009ihh}, or indirectly with ground-based Cherenkov telescopes, such as H.E.S.S. \cite{Hofmann:1999ew}, MAGIC \cite{Rico:2017euq}, and VERITAS \cite{Weekes:2001pd}. The observed photon flux exhibits a suppression at GeV energies relative to the predictions of electromagnetic cascades, which can be explained by considering the magnetic deflection of the produced electron-positron pairs by the intergalactic magnetic field (IGMF) \cite{aharonian2001tev,Neronov:2006lki,Neronov:2009gh, Neronov:2010gir,Taylor:2011bn,Vovk:2011aa}. Then, the secondary emission appears as a spatially extended halo around the source, reducing the observed gamma-ray flux. The absence of extended or delayed GeV-band emission from several extragalactic TeV sources has previously been used to place a lower bound on the strength of IGMF \cite{Fermi-LAT:2018jdy,MAGIC:2022piy,HESS:2023zwb,Vovk:2023qfk,Blunier:2025ddu,Vovk:2025rkr} in cosmic voids. Another possibility is that beam plasma instabilities induce energy loss of $e^{+}e^{-}$ pairs, which suppresses the IC cooling. The interaction between the relativistic blazar-induced pair beam and the intergalactic plasma can excite electrostatic waves, providing an additional pathway for energy dissipation that may compete with IC cooling. The efficiency and impact of these electrostatic instabilities continue to be actively explored through theoretical and numerical investigation \cite{Broderick:2011av, Miniati:2012ge, schlickeiser2012plasma, schlickeiser2013plasma, Sironi:2013qfa, Vafin:2018kox, AlvesBatista:2019ipr, Alawashra:2024fsz}. Previous works \cite{AlvesBatista:2019ipr, Saveliev:2013jda} have studied the growth of plasma instabilities in the IGM under various beam and environmental conditions, IGM temperature and density, and have shown that the energy losses due to plasma instabilities can reproduce the observed suppression in the GeV range of the gamma-ray spectra of blazars. In our earlier work \cite{Dey:2025wdj}, we investigated the impact of plasma instabilities on electromagnetic cascades using kinetic simulations and found that the resulting fractional energy loss is approximately $\lesssim 4\%$.

Constraints on long-coherence-length IGMFs reported in \cite{Neronov:2010gir, Tavecchio:2010ja} suggest a magnetic field strength $B\gtrsim 10^{-16}\text{~G}$, assuming that the currently observed TeV fluxes of selected blazars reflect their typical activity over $\sim 10^{6}$ yr. If this long-term activity assumption is relaxed and the sources are instead taken to be active only during the observational period, the bound weakens to $B\gtrsim 10^{-17}\text{~G}$ \cite{Dermer:2010mm, Taylor:2011bn, Vovk:2011aa}. Using 7.5 years of Fermi-LAT data \cite{atwood2009large}, the study reported by \cite{Fermi-LAT:2018jdy} derived the constraint of $B\gtrsim 3\times 10^{-16}\text{~G}$. However, it still relies on unverified assumptions about the decade-scale stability of TeV emission. Additional constraints come from searches for extended emission around Mrk 421 and Mrk 501 with MAGIC, which exclude $(0.4-1)\times 10^{-14}\text{~G}$ assuming persistent emission above 20 TeV \cite{MAGIC:2010goh}, and from H.E.S.S. observations that exclude IGMF strengths below $3\times 10^{-16}\text{~G}$ and up to $3\times 10^{-15}\text{~G}$ under similar assumptions \cite{HESS:2014kkl}. A study combining MAGIC and Fermi-LAT observations of the blazar 1ES 0229+200 derived a lower bound of $B \sim 1.8 \times 10^{-17}\text{~G}$ \cite{MAGIC:2022piy}. The LHAASO and Fermi-LAT observations of GRB 221009A reported $B\gtrsim 10^{-19}\text{~G}$ for $\lambda_{\text{c}}>1\text{~Mpc}$ \cite{Vovk:2023qfk}. Moreover, \cite{Burmeister:2025lgo} derived an improved constraint from the Fermi-LAT and LHAASO observations of the source GRB 221009A, excluding $B < 2.5 \times 10^{-17}~\text{G}$ at 95\% confidence level for $\lambda_{\text{c}}\gtrsim 1~\text{Mpc}$. This represents one of the weakest currently available limits. Another study by \cite{Webar:2025qbp} reports the detection of extended GeV emission around Mkn 501, interpreting it as evidence for an IGMF strength $\sim 1.5\times10^{-15}~\text{G}$ with a coherence length of $\sim 10~\text{kpc}$. Their analysis is based on searches for extended emission using the point-spread function (PSF) of the Fermi-LAT telescope. In addition, \cite{Keita:2026crk} predict that the upcoming Cherenkov Telescope Array Observatory (CTAO) observations of bright events such as GRB 221009A and GRB 190114C could probe IGMF strength as strong as $B \sim 10^{-15}\text{~G}$.

When a TeV blazar emits gamma rays continuously over a period much longer than the typical cascade-photon delay time, the delays are effectively ``washed out,'' and the observer measures the total emission spectrum, including all components. Since 1ES 0229+200 shows no evidence of rapid variability (see \cite{MAGIC:2022piy} for reference), we concentrate our analysis on the angular extension of the emission, rather than on time-delay methods as used in \cite{MAGIC:2022piy, Vovk:2023qfk}. 

In this study, we introduce a two-parameter energy-loss term, following an approach similar to \cite{AlvesBatista:2019ipr}, to describe the cooling of electron–positron pairs due to plasma instabilities. We simulate the propagation of VHE gamma rays using the Monte Carlo framework \texttt{CRPropa 3.2} \cite{AlvesBatista:2022vem}, while accounting for all relevant electromagnetic processes, the additional energy-loss term associated with plasma instabilities, and the presence of an IGMF. Further, we investigate how plasma instabilities affect the measurement of the IGMF using Fermi-LAT observations of the blazar 1ES 0229+200. The paper is organized as follows: Section \ref{sec:plasma-inst} introduces the parametric model of plasma instability. Section \ref{sec:sim} describes the cascade simulation configuration, including the jet emission scheme, observer setup, IGMF configuration, and other relevant interactions for cascade production. Section \ref{sec:output} presents the output analysis methods, including cascade spectrum construction, field-of-view analysis, and test statistics. In Section \ref{sec:sim-results} we present the simulation results for electromagnetic cascades with IGMF only, with plasma instability only, and with both IGMF and plasma instability. Section \ref{sec:Lower-lim-with-inst} discusses the impact of plasma instabilities on IGMF constraints. Finally, Section \ref{sec:disscussion} offers a discussion and conclusion.

\section{Parametric model of plasma instability}\label{sec:plasma-inst}
Low-energy background photon fields, such as the EBL, are optically thick to high-energy $\gamma$-rays, leading to the production of relativistic, low-density pair beams. As these beams propagate through the intergalactic medium (IGM), fluctuations in their momentum distribution can perturb the background plasma and thereby trigger different plasma instability modes. In such a situation, Langmuir waves excited in the background plasma can resonate with oscillations in the beam, leading to the exponential growth of electrostatic fluctuations \cite{bret2010multidimensional}. The characteristic spatial scale for the beam--plasma instability growth is given by the plasma skin depth of the IGM, which is defined as $\lambda_{\text{s}} = 2\pi c/\omega_{p}$,
where the plasma frequency is given by, $\omega_{p} = (4\pi e^{2} n_{\text{IGM}}/m_{e})^{1/2}$ and $n_{\text{IGM}}$ represents the plasma number density in the IGM. In physical terms, plasma effects become important when the number of beam pairs contained within a volume of radius $\lambda_{\text{s}}$ is much greater than unity \cite{Broderick:2011av}, i.e.,
\begin{equation}
    \lambda_{\text{s}}^3 n_{\text{b}}\gg 1,
    \label{criteria-skin-depth}
\end{equation}
where $n_{\text{b}}$ is the number density of pair beam. For standard astrophysical scenarios $n_{\text{b}}\sim 10^{-22}$ cm$^{-3}$ and $n_{\text{IGM}}\sim 10^{-7}$ cm$^{-3}$ \cite{Broderick:2011av}, the condition in Eq. \eqref{criteria-skin-depth} is satisfied, indicating that beam–plasma instabilities can be relevant in astrophysical scenarios. Astrophysical pair beams, such as those produced by the interaction between TeV photons and background photon fields, are characterized by an intrinsically anisotropic momentum distribution \cite{Broderick:2011av}. The anisotropy arises from the fact that the longitudinal momentum (i.e., the momentum projection along the beam direction) is considerably larger, as the pairs inherit the high energy of the parent TeV photon, and the transverse momentum (i.e., the component perpendicular to the beam direction) is much smaller. This corresponds to a small angular spread, since pair production tends to conserve the direction of the incident photon, suppressing electromagnetic instabilities and making electrostatic instabilities the dominant ones \cite{Dey:2025wdj, Vafin:2018kox}. Among the various electrostatic instability modes, the oblique and modulation instabilities are the most relevant in this context, as they are the fastest-growing resonant modes \cite{Bret:2009fy, Vafin:2018kox, Dey:2025wdj}. The oblique instability arises from the coupling between beam-driven electrostatic oscillations and background ion-density perturbations in a plasma. The plasma wave propagates at an angle between $0<\theta<\pi/2$, relative to the beam direction, allowing both longitudinal and transverse wave components to contribute to the coupling. The oblique instability is basically the same as the two-stream-like instability, except that the wave vector is not parallel with the beam velocity, but at an angle to the beam direction. The modulation instability, on the other hand, can be regarded as the oscillating two-stream-like instability, except that the interacting ``streams'' are the forward and backward Langmuir sidebands instead of two particle beams.\\
The propagation of the electron-positron pairs is affected by both IC scattering with CMB photons \cite{Penzias:1965wn} and beam--plasma instabilities, as discussed above. Therefore, the relevant length scales are the instability cooling length, $\lambda_{0}$, (i.e., the typical distance an electron/positron travels before losing significant energy to plasma instability processes) and the IC interaction length, $\lambda_{\text{IC}}$, (i.e., the average distance the electron travels before undergoing one IC scattering with a CMB photon). If $\lambda_{0}$ is comparable or smaller than $\lambda_{\text{IC}}$, plasma processes dominate particle cooling, suppressing the production of secondary GeV photons. \\
We define a two-parameter energy-loss term (following the approach described in \cite{Castro:2024ooo}) to parameterize the plasma instability cooling of electron–positron pairs. Therefore, the energy-loss rate for electrons with energy $E_e$ due to plasma instabilities is given by
\begin{equation}\label{frac-loss-eqn}
    E_{e}^{-1}\left(\frac{dE_{e}}{dt}\right)= c \lambda_{0}^{-1} \left(\frac{E_e}{\tilde{E_e}}\right)^{-\alpha}~\text{sec}^{-1}.
\end{equation}
We perform a parametric analysis using two parameters, the instability length scale, $\lambda_{0}$, and the power-law index, $\alpha$, while keeping the normalization energy fixed at $\tilde{E_{e}} = 1 ~\text{TeV}$. In this work, we do not resolve the coupling between individual electrons (or positrons) and plasma waves. Instead, we consider the single-particle level by defining a parametric energy-loss rate in Eq. \eqref{frac-loss-eqn}. We use the publicly available Monte Carlo framework \texttt{CRPropa 3.2} \cite{AlvesBatista:2022vem}\footnote{\href{https://crpropa.desy.de/}{https://crpropa.desy.de/}} to simulate the transport of electromagnetic particles and the development of electromagnetic cascades. A similar approach has been adopted previously in \cite{AlvesBatista:2019ipr}, where the plasma instability energy-loss rates were obtained using both analytical studies \cite{Broderick:2011av, Miniati:2012ge, schlickeiser2012plasma, schlickeiser2013plasma} and kinetic simulation results \cite{Sironi:2013qfa, Vafin:2018kox}. However, the assumption in \cite{AlvesBatista:2019ipr} that the electron energy-loss timescale, $\tau_{\text{inst}}$, is equal to the instability growth timescale, $\tau_{\text{gr}}$, differs from the treatment adopted in \cite{Sironi:2013qfa, Vafin:2018kox}. In particular, equating these two timescales corresponds to assuming that the beam energy loss proceeds on the same timescale as the instability growth. In this work, we generalize the parametrization of the energy-loss term according to \cite{Castro:2024ooo} and estimate the fractional energy loss associated with the instability, and therefore do not rely on the previous assumption. According to previous studies \cite{Broderick:2011av, Miniati:2012ge, schlickeiser2012plasma, schlickeiser2013plasma, Sironi:2013qfa, Vafin:2018kox}, the energy dependence for both cold- and warm-plasma beams ranges from a power-law index range of $-2.0 \leq \alpha \leq 1.6$. For warm plasma beams where the oblique instability dominates, typical values of $\alpha$ lie in the negative range $-1 \leq \alpha \leq 0$. In this work, we adopt this energy dependence of the instability energy-loss term over the same range of $\alpha$ in our simulations, consistent with the warm-plasma beam regime in which the oblique instability is the dominant mode. In realistic scenarios, the energy-loss length scale is always greater than the instability growth length scale. For example, the minimum instability growth length is approximately in the range $\lambda_{\text{gr}}\approx 1.43 - 2.85$ pc for fiducial astrophysical parameters, with beam densities $n_{\text{b}}\sim [0.5-1]\times10^{-23}$ cm$^{-3}$, $n_{\text{IGM}}\sim 10^{-7}$ cm$^{-3}$ and the pair beam with Lorentz factor $\gamma_\text{beam}\sim 10^{5} - 10^{6}$ \cite{Alawashra:2022all}. The energy-loss length scale, accounting for the nonlinear evolution of the electrostatic wave as described in \cite{Vafin:2018kox, Alawashra:2022all}, is estimated to be $\lambda_{0}\simeq 5\times 10^{4}\,\lambda_{\text{gr}}= [0.7-1.4]\times10^2$ kpc. Therefore, we choose the value of $\lambda_{0}$ within the range $100 - 140~\text{kpc}$, to evaluate the fractional energy loss due to plasma instabilities over a single IC scattering length, $\lambda_{\text{IC}}$. We then determine the best-fit values of $\lambda_{0}$ and $\alpha$ considering the case where the development of the cascades is only affected by plasma instabilities, reproducing the results of \cite{Castro:2024ooo}. Our treatment of the energy loss of the pair beam, representing the beam--plasma instability, is based on the following set of assumptions:
\begin{enumerate}
    \item We assume spatially uniform energy losses, such that all electrons lose their energy irrespective of their distance from the source.
    \item Although plasma instability eventually saturates after the linear growth due to non-linear effects caused by the instability feedback, we assume that our implementation remains valid within the regime where the instability undergoes growth between two different IC scatterings. That means, the instability saturation time, $\tau_{\text{sat}}$, is longer than the IC interaction time, $\tau_{\text{IC}}$, i.e., $\tau_{\text{IC}}<\tau_{\text{sat}}$, ensuing that instability remains in its growth phase between two different IC scatterings. For electrons with energies in the range $10^{-3}\leq E_{e}/\text{TeV} \leq 10^{2}$, IC interaction length is nearly independent of energy, with $\lambda_{\text{IC}} \sim 1.2$ kpc, corresponding to an IC interaction time of $\tau_{\text{IC}}\sim 1.23\times10^{11}$ sec \cite{Neronov:2010gir}. According to \cite{Alawashra:2024fsz}, the instability saturation time is approximately $\tau_{\text{sat}}\sim 5\times 10^{13}$ sec, which justifies our assumption.
\end{enumerate} 

\section{Cascade simulation configuration}\label{sec:sim}
This section describes the numerical modeling considered in this work for the blazar injection and propagation of the secondary particles produced in the electromagnetic cascades.

\subsection{Jet emission scheme and observer setup}
In this work, we model the high-energy photon emission geometry of the blazar 1ES 0229+200 with \texttt{CRPropa 3.2}. In particular, we consider a von Mises-Fisher distribution \cite{Fisher1953} for the distribution of the initial photon momentum at the source, which corresponds to the generalization of a normal distribution on a sphere. The implementation of the von Mises-Fisher distribution in \texttt{CRPropa 3.2} can be found in \cite{Jasche:2019sog,AlvesBatista:2022vem}. 

Given the unit vector $\hat{\boldsymbol{n}}$, the von Mises-Fisher distribution of $\hat{\boldsymbol{n}}$ on the unit sphere is given by
\begin{equation}
    \label{vMF_distribution}
    \Pi_\text{vMF}(\hat{\boldsymbol{n}};\hat{\boldsymbol{\mu}},\kappa) = \dfrac{\kappa}{2\pi\left(1-e^{-2\kappa}\right)}\exp\left[\kappa\left(\hat{\boldsymbol{n}}\cdot \hat{\boldsymbol{\mu}}-1\right)\right] \, ,
\end{equation}
where $\hat{\boldsymbol{\mu}}$ is a unit vector representing the mean injection direction and $\kappa>0$ is the concentration parameter, which parameterizes the width of the distribution. We can see that for $k\ll1$, the von Mises-Fisher distribution converges to the isotropic distribution $1/4\pi$. The concentration parameter $\kappa$ can be used to parameterize the opening angle $\delta_\text{jet}$ of the jet. Assuming that a certain fraction $P$ of emitted photons is contained within the jet axis and the opening angle $\delta_\text{jet}$, from Equation \eqref{vMF_distribution} we obtain that
\begin{equation}
    \label{vMF_fraction}
    P=\dfrac{1-e^{\kappa(\cos\delta_\text{jet}-1)}}{1-e^{-2\kappa}} \, .
\end{equation}
For $\kappa\gg1$, the concentration parameter is given by
\begin{equation}
    \kappa\simeq\dfrac{\ln\left(1-P\right)}{\cos\delta_\text{jet}-1} \, .
\end{equation}
We choose $\kappa = 196$ because it produces an angular concentration for which approximately 95\% of the injected photons fall within the jet opening angle $\delta_{\text{jet}} = 10^\circ$. In CRPropa, observers are implemented as surfaces, such as spheres, that register particles upon crossing and record both their propagated-state properties (energy, position, momentum, particle type) and their initial source parameters in the output file. In this work, we construct an observer consisting of a spherical surface of radius $d=585\text{~Mpc}$ centered on the source.

\subsection{Interactions for cascade development}
TeV photons emitted by extragalactic sources propagate through cosmic radiation fields, such as the CMB and the EBL, leading to the development of electromagnetic cascades. The development of the cascade begins when the high-energy photons interact with background photons and produce electron–positron pairs. The resulting charged particles inherit most of the initial photon energy and continue to propagate predominantly in the forward direction, where they may either lose energy through plasma instabilities or experience deflections by the weak IGMF, and subsequently upscatter background photons through IC scatterings. The newly produced secondary photons can again undergo pair production, and the sequence repeats until the energies of the particles fall below the relevant interaction thresholds. We include the interaction rates for pair production and IC scattering on the CMB and the infrared background (IRB) photons from the \texttt{Franceschini08} model \cite{franceschini2008extragalactic}, which are already available in CRPropa. We also incorporate double- and triple-pair production processes. However, their contributions are negligible for photon energies below $\sim 100\text{~TeV}$.

The magnetic deflection of the electron-positron pairs produced in the cascade is taken into account by introducing a homogeneous turbulent magnetic field that follows a Kolmogorov power spectrum, concentrated in a single mode at the wavenumber $\lambda_{\text{c}}^{-1}$, characterized by the spectral index $\mathcal{K} = 5/3$. The magnetic field coherence length, $\lambda_{\text{c}}$, is determined by the maximum and minimum turbulence scales, $\lambda_{\max}$ and $\lambda_{\min}$, respectively, and is expressed as \cite{Dundovic:2017vsz}
\begin{equation}
\lambda_{\text{c}} = \frac{1}{2\lambda_{\max}} \cdot \frac{\mathcal{K} - 1}{\mathcal{K}} \cdot \frac{1 - (\lambda_{\min}/\lambda_{\max})^\mathcal{K}}{1 - (\lambda_{\min}/\lambda_{\max})^{\mathcal{K} - 1}},
\label{eq:coherence-length}
\end{equation}
Since $\lambda_{\max}$ and $\lambda_{\min}$ serve as direct inputs to the simulation, this relation must be solved numerically to obtain a target coherence length $\lambda_{\text{c}}$. The discretized nature of the magnetic field grid in CRPropa, however, imposes physical limits on the turbulence scales: for a grid with $N$ number of cells of size $\Delta x$, the turbulent field is fully resolved only if $\lambda_{\max} < N\Delta x$ and $\lambda_{\min} > 2\Delta x$. For our setup with $N = 256$, the parameters $\lambda_{\min}$ and $\lambda_{\max}$ are selected in accordance with these grid constraints to ensure a physically consistent realization of the turbulent field.

\section{Predictions of the \texorpdfstring{$\gamma-$}{gamma-}ray spectra}\label{sec:output}
We use reweighting methods to reproduce the observed emission spectra of blazars accurately. We calculate a cascade signal function $G(E_{0}, E;\lambda_{0}, \alpha; \theta_{\text{FoV}}, \lambda_{\text{c}})$ that relates the primary photon energy $E_0$ to the observed photon energy $E$ for different values of IGMF coherence length, observational field-of-view, and the parameters representing the plasma instability scenario. Each combination of IGMF strength and coherence length yields a distinct cascade signal function for a specific plasma instability case. Furthermore, during the cascade development, multiple secondary photons ($\sim$GeV) are produced, meaning that an observed photon with energy $E$ originates from one of several photons generated by an injected photon with energy $E_{0}$. We simulate the resulting photon spectrum for a fixed set of plasma instability parameters and determine the corresponding IGMF strength and coherence length that provide the best fit to the observational data for a particular field-of-view observation. The propagation of injected and pair-produced electrons and positrons is modeled in three dimensions in order to account for their deflection by the IGMF. 
\subsection{Cascade spectrum construction}
The H.E.S.S. and VERITAS data do not show any significant spectral variability of 1ES 0229+200 above $100$ GeV \cite{MAGIC:2022piy}. We consider an intrinsic source spectrum that follows a power-law function with an exponential cutoff
\begin{equation}
    J_{0}(E_{0}) = A \left(\frac{E_{0}}{\text{TeV}}\right)^{-\beta} \text{exp}\left(-\frac{E_{0}}{E_{\text{cut}}}\right),
    \label{eq:power-law}
\end{equation}
where $E_{0}$ is the energy of the primary photon, $\beta$ represents the source spectral index, $E_{\text{cut}}$ is the high-energy cutoff, and $A$ is a normalization constant.
The propagated photon spectrum $J(E; \lambda_{0}, \alpha; \lambda_{\text{c}}, B, \theta_{\text{FoV}})$ observed at Earth can be obtained by reconstructing the cascade signal $G(E_{0}, E;\lambda_{0}, \alpha; \lambda_{\text{c}}, B, \theta_{\text{FoV}})$ with the initial injection spectrum $J_{0}(E_{0})$, expressed as \cite{MAGIC:2022piy},
\begin{equation}
\begin{split}
        J(E; \lambda_{0}, \alpha; & \lambda_{\text{c}}, B, \theta_{\text{FoV}}) = \\
        &\int_{E_{0}\geq E} G(E_{0}, E;\lambda_{0}, \alpha; \lambda_{\text{c}}, B, \theta_{\text{FoV}})\cdot J_{0}(E_{0})~dE_{0},
    \label{eq:reconstruction-spectrum}
\end{split}
\end{equation}
We simulate the propagation of $10^{4}$ primary photons with energies in the range $10^{-3}\leq E_{0}/\text{TeV} \leq 10^{2}$, with a propagation step size of $\Delta x \simeq 10^{14}$ cm, up to a comoving distance of $d = 585$ Mpc.

\subsection{FoV analysis method}  
\begin{figure*}
   \centering
   \includegraphics[width=0.8\textwidth]{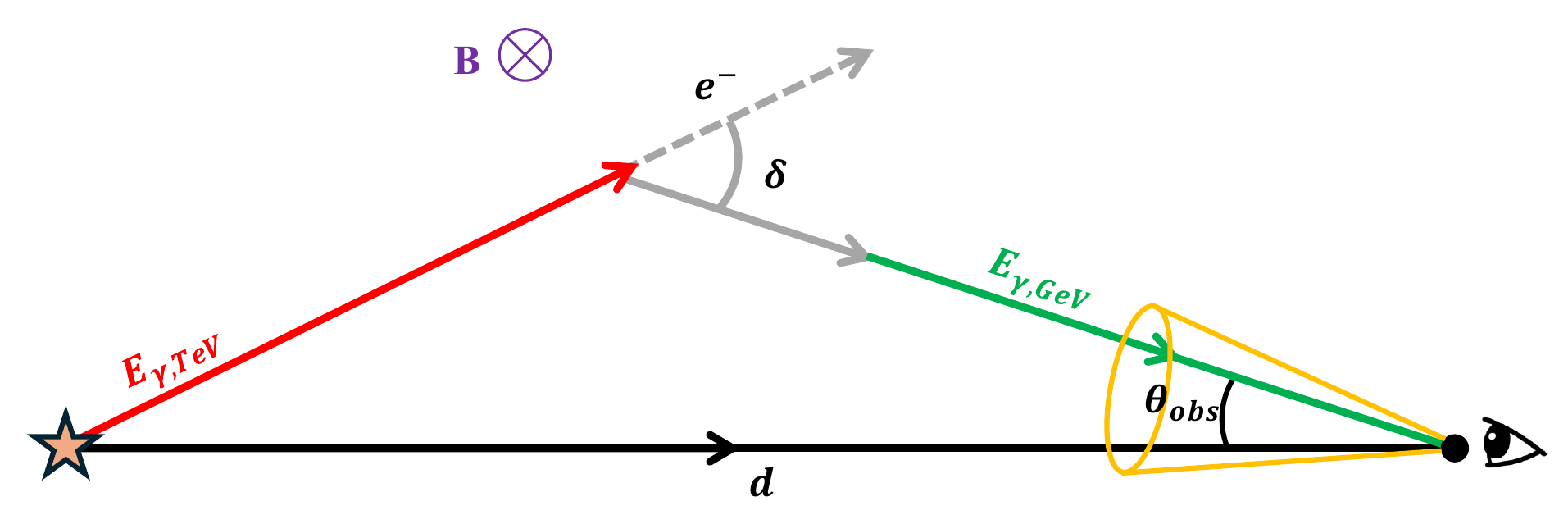}
      \caption{Schematic representation of the production of secondary GeV photons from the deflection of charged electrons (or positrons) by IGMF. The solid gray line indicates the deflected electron due to the IGMF, and the dashed gray line shows the original direction of the electron without any deflection. The yellow region represents the field of view of the observer, defined by the angle $\theta_{\text{FoV}}$.}
              \label{fig:schematic}%
\end{figure*}

When a TeV blazar emits gamma rays continuously over a period much longer than the typical delay time of cascade photons, i.e., $\tau_{\text{emission}}>\tau_{\text{delay}}$, the time-delay effects are washed out, and the observation is not affected by the propagation delay anymore. Since the blazar 1ES 0229+200 exhibits no significant rapid variability, we focus our analysis on the angular extension of the observed emission rather than the time-delay approach adopted in \cite{MAGIC:2022piy, Vovk:2023qfk}. The exchange of momentum during the interaction of a primary TeV photon with an EBL photon to produce an electron-positron pair is governed by relativistic kinematics. In the IGM rest frame (lab frame), this pair production process takes place when a primary TeV photon, with four-momentum $P_{0} = E_{0}(1, \mathbf{\hat{p}_{0}})$, interacts predominantly with a low-energy EBL photon of four-momentum $P_{1} = E_{\text{EBL}}(1, \mathbf{\hat{p}_{1}})$, where $\mathbf{\hat{p}_{0}}$ and $\mathbf{\hat{p}_{1}}$ are the respective unit vectors of momentum. The four-momentum of the produced electron and positron is described by $P_{\pm} = (E_{e\pm}, \mathbf{q_{\pm}})$ \cite{Schlickeiser:2012hn}. It is important to note that the angle between the momentum of the incident photon and that of the pair-produced electron/positron (also called the initial beam opening angle), produced with Lorentz factors $\Gamma\sim10^{5}-10^7$, is approximately given by $\sim \Gamma^{-1}$. Hence, this angle is negligibly small during the pair production process \cite{Perry:2021rgv}. 
Before the electrons (and positrons) primarily up-scatter CMB photons to GeV energies via IC scattering, they are deflected by the IGMF or simultaneously experience energy loss due to plasma instabilities. The deflection angle of the electron (or positrons) by IGMF, with respect to the propagation direction of the primary TeV photon, is given by $\delta = \text{cos}^{-1}(\mathbf{\hat{p}_{0}\cdot\hat{q}_{\pm}^{\prime}})$. The unit vector of momentum of an electron (or positron) after deflection, $\mathbf{\hat{q}_{\pm}^{\prime}}$, and $\mathbf{\hat{p}_{0}}$, is directly obtained from the simulation output. 
The resulting secondary photons produced from IC scattering with energy $E_{\gamma}$, are observed within an angle, $\theta_{\text{obs}}$, which depends on the $\gamma$-ray mean free path, $\lambda_{\gamma}$, the comoving distance between the source and observer, $d$, and the deflection angle of the electron, $\delta$ \cite{Neronov:2009gh},
\begin{equation}
  \text{sin}\left(\theta_{\text{obs}}\right) = \frac{\lambda_{\gamma}\left(E_{0}\right)}{d} \text{sin }\delta,
    \label{eq:extended-angle}
\end{equation}
The mean free path of a primary photon with energy $E_{0}$ for pair production with EBL photons is approximately $\lambda_{\gamma} \approx 44.36\times(E_{0}/10\text{~TeV})^{-1}$ Mpc for redshifts of $z<1$ \cite{Kneiske:2003tx, Neronov:2009gh, Alawashra:2024fsz}. Using Eq.~(\ref{eq:extended-angle}), we estimate $\theta_{\text{obs}}$ and then define the observer’s field of view angle $\theta_{\text{FoV}}$, as an independent parameter to evaluate the cascade-induced ``glow'' within this angular region. A schematic representation of the electromagnetic cascade formation due to the deflection of charged electrons (or positrons) by IGMF is shown in Fig.~\ref{fig:schematic}. The angular resolution, or PSF, of Fermi-LAT is energy-dependent, with a 68\% containment angle of approximately $0.7^{\circ}$ at $1$ GeV and $0.08^{\circ}$ at $1$ TeV \cite{atwood2009large, Fermi-LAT:2013jgq}\footnote{\url{https://www.slac.stanford.edu/exp/glast/groups/canda/lat_Performance.htm}}. We select representative values of $\theta_{\text{FoV}}=1.0^{\circ}$ and $4.5^{\circ}$, consistent with the observational constraint $\theta_{\text{psf}} < \theta_{\text{obs}} < \theta_{\text{FoV}}$ \cite{AlvesBatista:2021sln}. We simulate the photon spectrum for different configurations of magnetic field strengths and $\lambda_{\text{c}}$ values within these fields of view, incorporating the plasma instability cooling.

\subsection{Test statistic}\label{subsec:test-stat}
We define the $\chi^{2}$ test statistic to determine the deviation of the simulated spectrum from the observation as
\begin{equation}
\begin{split}
    &\chi^{2}(A, \beta, E_{\text{cut}}; \lambda_{0}, \alpha, \lambda_{\text{c}}, B, \theta_{\text{FoV}})\\ &= \sum_i\left(\frac{J_{\text{data}}(E_{i})-J_{\text{sim}}(E_{i};A, \beta, E_{\text{cut}}; \lambda_{0}, \alpha, \lambda_{\text{c}}, B, \theta_{\text{FoV}})}{\sigma_{\text{data},i}}\right)^{2},
    \end{split}
    \label{eq:chisq}
\end{equation}
where $J_{\text{data}}(E_{i})$ and $J_{\text{sim}}(E_{i};A,\beta, E_{\text{cut}}; \lambda_{0}, \alpha, \lambda_{\text{c}}, B, \theta_{\text{FoV}})$ represent the observed and propagated photon spectra at Earth, respectively, evaluated at the same energy $E_{i}$. The summation is carried out over the observed energy bins. The uncertainty associated with each observed data point is denoted by $\sigma_{\text{data},i}$. The $\chi^{2}$ analysis is performed within field-of-view angles $\theta_{\text{FoV}}=1.0^{\circ}$ and $4.5^{\circ}$, first for the case without plasma instability and then including the best-fit plasma instability parameters following \cite{Castro:2024ooo}. We normalize the value of $\chi^{2}$ by the number of degrees of freedom, $n_{\text{dof}}$, defined as the difference between the total number of data points across the energy bins used in Eq.~(\ref{eq:chisq}) and the number of fitted parameters. We obtain the best-fit intrinsic spectral parameters $A, \beta, E_{\text{cut}}$ separately for each combination of $B$ and $\lambda_{\text{c}}$, with 90\% confidence level. The parameter space is scanned over $\beta = 1.0, \cdots,2.0$ with a step size $\Delta \beta = 0.01$, $E_{\text{cut}} = 1.0, \cdots,15.0~\text{TeV}$ with $\Delta E_{\text{cut}} =0.1~\text{TeV}$, and $\text{log}_{10}A = -24.000, \cdots,-23.200$ eV$^{-1}$cm$^{-2}$sec$^{-1}$ with $\Delta \text{log}_{10}A =0.001$ eV$^{-1}$cm$^{-2}$sec$^{-1}$. Using the corresponding best-fit values of $A, \beta, E_{\text{cut}}$, we then calculate $\chi^2_{\min} 
= \min_{\lambda_{\text{c}}} [ \chi^2_{\min}(\lambda_{\text{c}}; B) ]
= \min_{\lambda_{\text{c}}} ( \min_{A, \beta,E_{\text{cut}}} 
\chi^2(A, \beta, E_{\text{cut}};\lambda_{\text{c}}; B) )$. In the absence of an IGMF, considering only plasma-instability cooling, we explore the parameter space over $\alpha = [-1.0,-0.5, 0.0]$ and $\lambda_{0}\in[100,140]~\text{kpc}$ with a step size $\Delta\lambda_{0}=20~\text{kpc}$. We obtain the best-fit intrinsic spectral parameters $A, \beta, E_{\text{cut}}$ separately for each combination of $\lambda_{0}$ and $\alpha$, at the 90\% confidence level. The scan over the intrinsic spectral parameters $A$, $\beta$, and $E_{\text{cut}}$ is performed using the same parameter ranges and step sizes as described before. For the plasma-instability-only case, the minimum chi-square is obtained as $\chi^{2}_{\text{min}} = \min_{\lambda_{0}, \alpha} [ \chi^2_{\min}(A, \beta, E_{\text{cut}};\lambda_{0}, \alpha) ]$. The fit is performed to the observed GeV data, as cascade emission is predominant in this energy range, while the TeV component of the spectrum remains unaffected. The total $n_{\text{dof}}=2$ when the GeV-band data from Fermi-LAT are considered. We apply a selection criterion requiring the minimum $\chi^{2}_{\text{min}}$ value to fall below the critical value of the $\chi^{2}$ distribution corresponding to a $2\sigma$ confidence level. In this case, the standard $2\sigma$ confidence interval is defined by $\Delta \chi^{2} =\chi^{2} - \chi^{2}_{\text{min}}\leq 8$. The normalized value corresponds to $\Delta \chi^{2} /n_{\text{dof}} \leq 4$, which is the standard normalized threshold for $n_{\text{dof}}=2$ for a $2\sigma$ deviation. 

\section{Simulation results}\label{sec:sim-results}
In this section, we discuss the propagated photon spectrum of the 1ES 0229+200 blazar under different propagation scenarios. We first consider the case without plasma instabilities, in which the pairs experience only magnetic deflection by the IGMF. We then present the propagated photon spectrum, accounting for both the effects of plasma instabilities and IGMF. We determine the combination of IGMF and plasma instability parameters that result in the minimum test statistic $\chi^{2}(A, \beta, E_{\text{cut}}; \lambda_{0}, \alpha; \lambda_{\text{c}}, B, \theta_{\text{FoV}})$, according to Eq.~(\ref{eq:chisq}).

\subsection{Electromagnetic cascade with IGMF only}\label{subsec:cascade-with-igmf-only}
\begin{figure*}[!ht]
   \centering
   \includegraphics[width=0.49\textwidth]{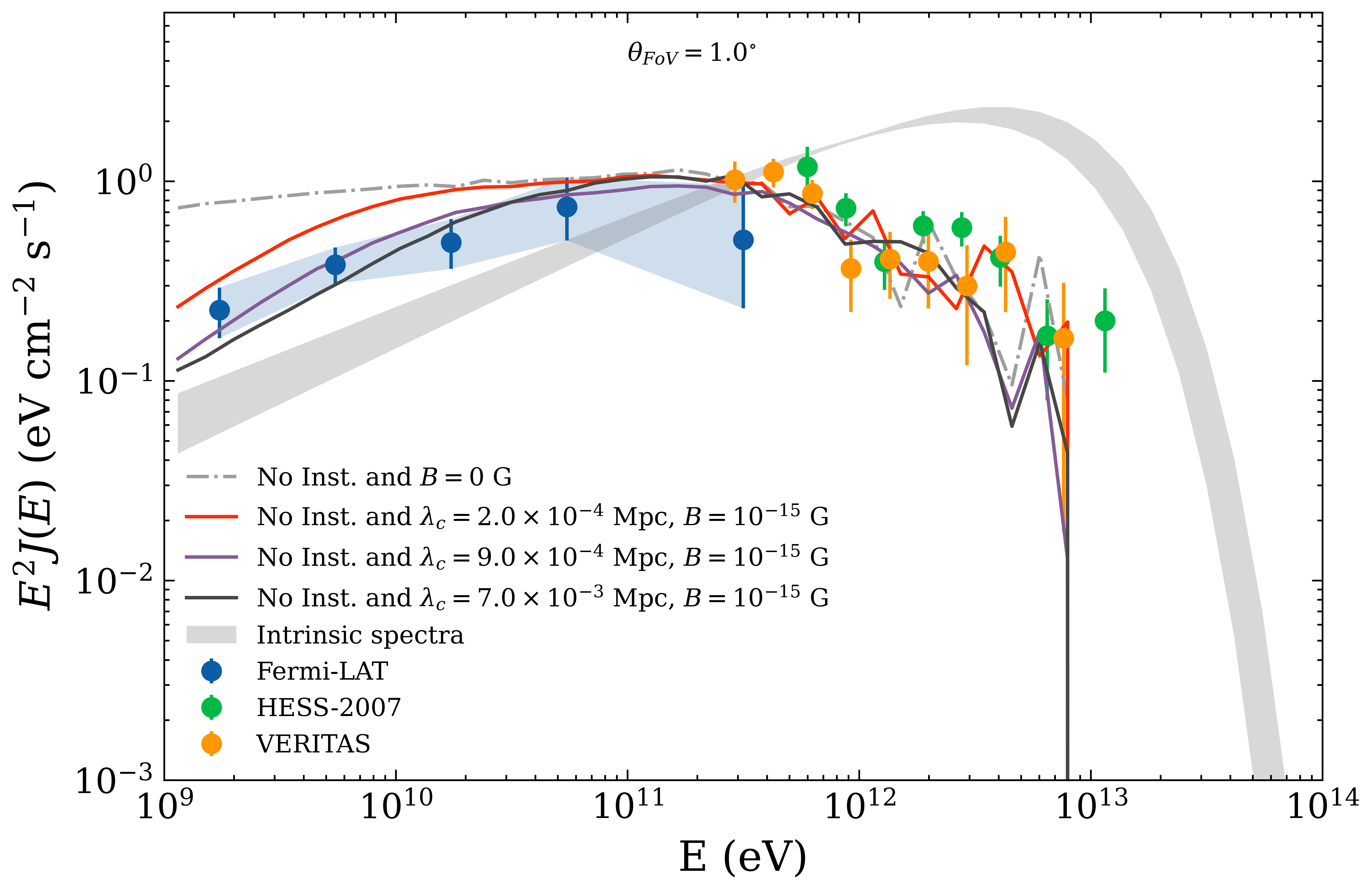}
   \includegraphics[width=0.49\textwidth]{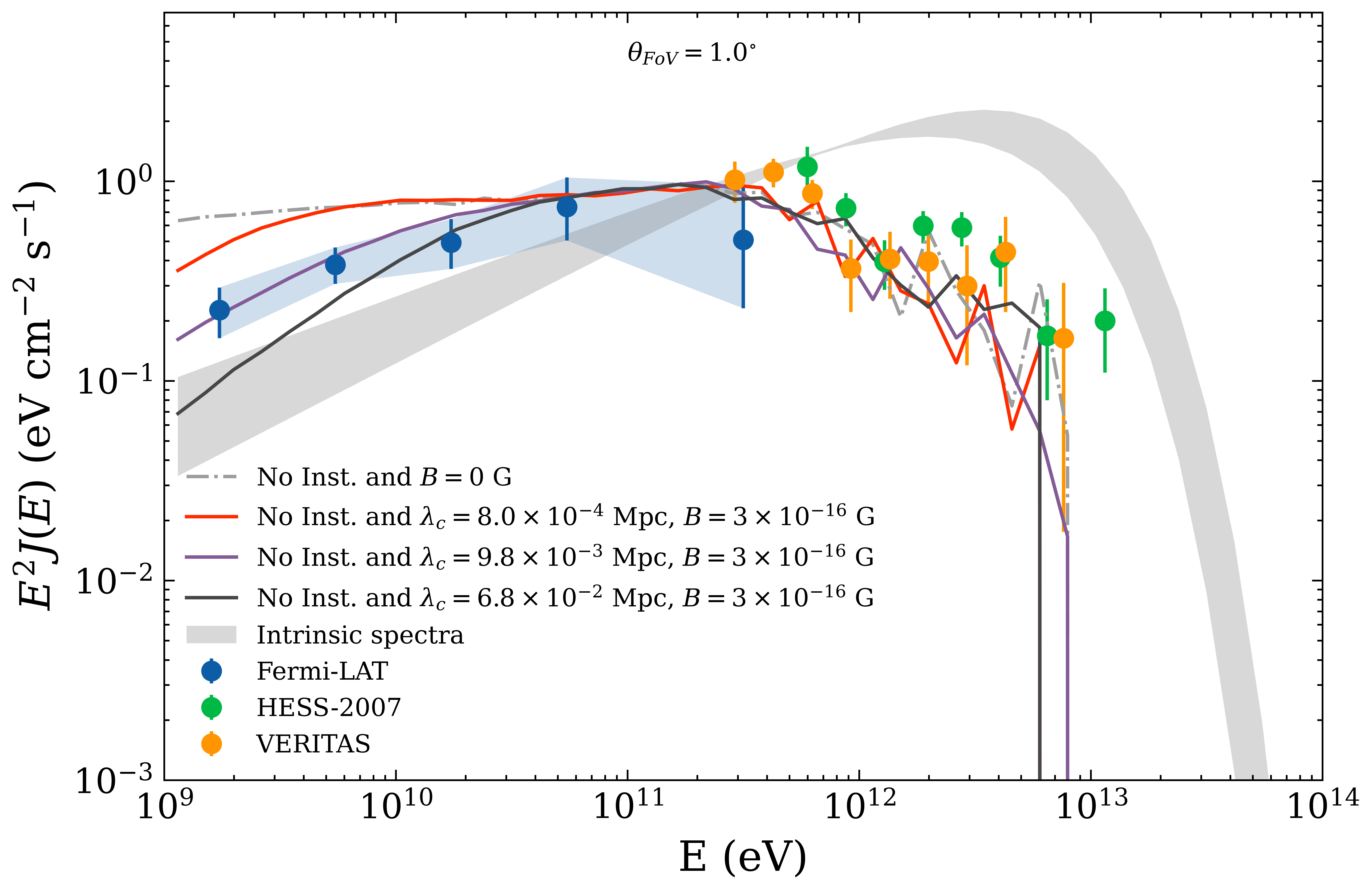}
   \includegraphics[width=0.49\textwidth]{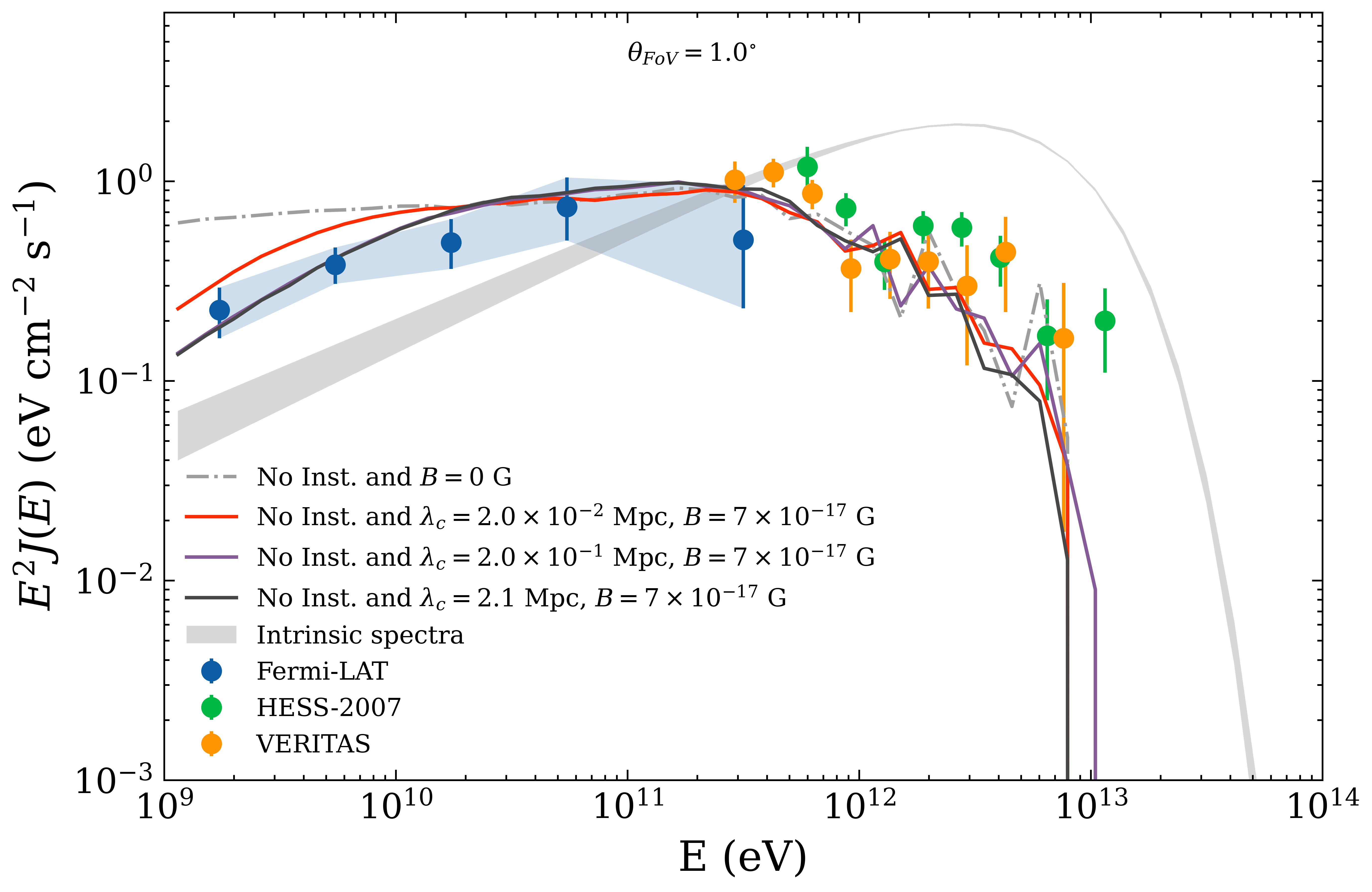}
      \caption{The energy spectrum of 1ES 0229+200 in $10^{-3}\leq E/\text{TeV} \leq 10^{2}$. The gray dash-dotted curve shows the propagated photon spectrum without any contribution of the plasma instability cooling (Inst. stands for Instability in the plot legends) and IGMF. The colored solid curves represent the propagated photon spectrum for different IGMF strengths and coherence lengths, without plasma instability cooling. The gray band represents the unattenuated intrinsic spectra, and the corresponding injection parameters are listed in Table~\ref{tab:chi-sq2-only-mag}. The blue data points represent the Fermi-LAT \cite{MAGIC:2022piy}, the green data points show the H.E.S.S. \cite{HESS:2007xak}, and the orange data points show the VERITAS \cite{Aliu:2013pya} spectrum. We investigate the energy spectrum, which is consistent with the Fermi-LAT, H.E.S.S., and VERITAS observational data in each plot. The energy spectrum is obtained for a fixed $\theta_{\text{FoV}}=1.0^{\circ}$. Different panels correspond to different field strengths, as indicated in the legends. The $\chi^{2}/n_{\text{dof}}$ test statistic for the GeV-band data from Fermi-LAT ($n_{\text{dof}}=2$) is listed in Table~\ref{tab:chi-sq2-only-mag}.}
              \label{fig:theta-1-0-only-mag}%
\end{figure*}

\begin{figure*}[!ht]
   \centering
   \includegraphics[width=0.49\textwidth]{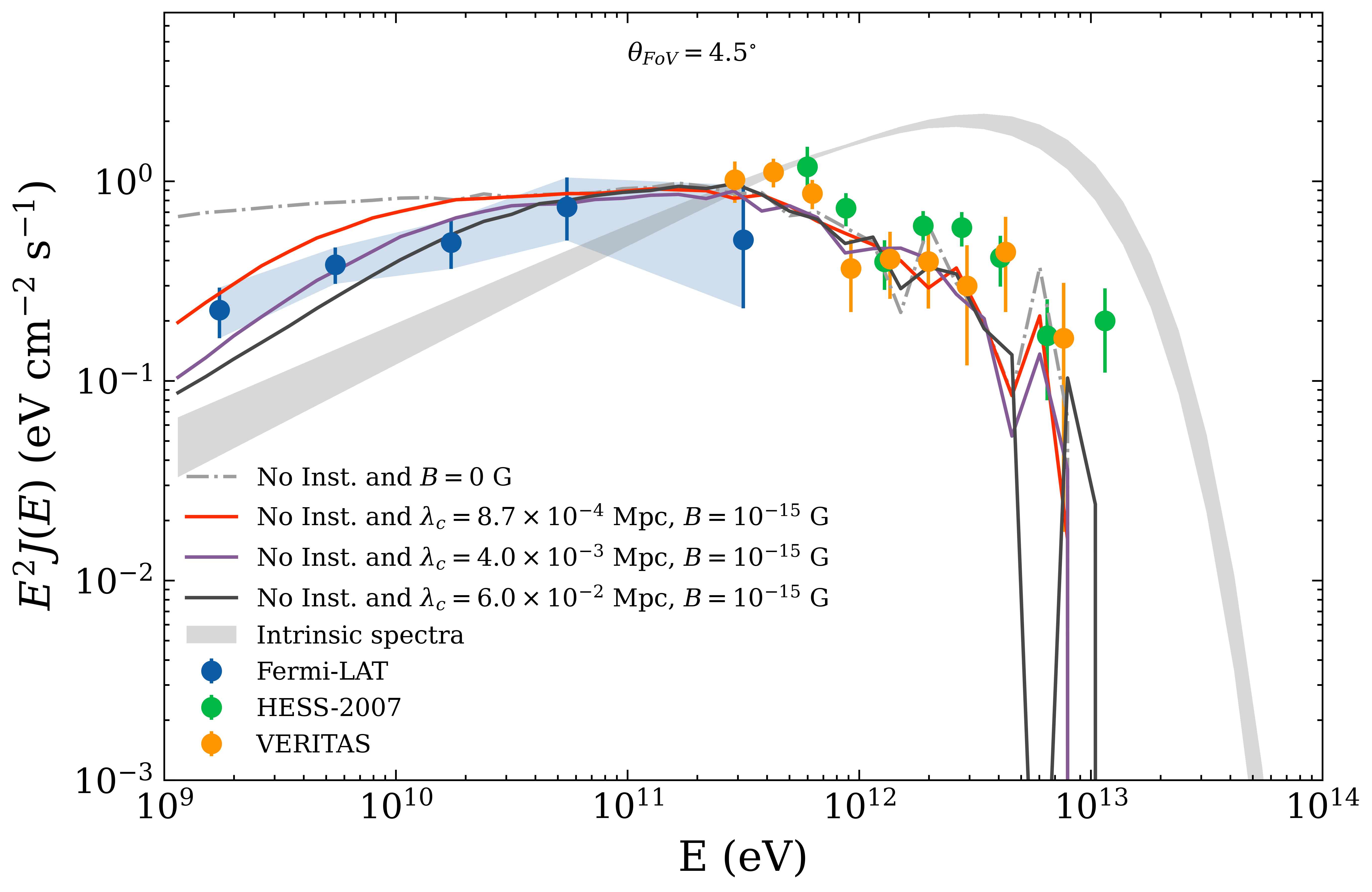}
   \includegraphics[width=0.49\textwidth]{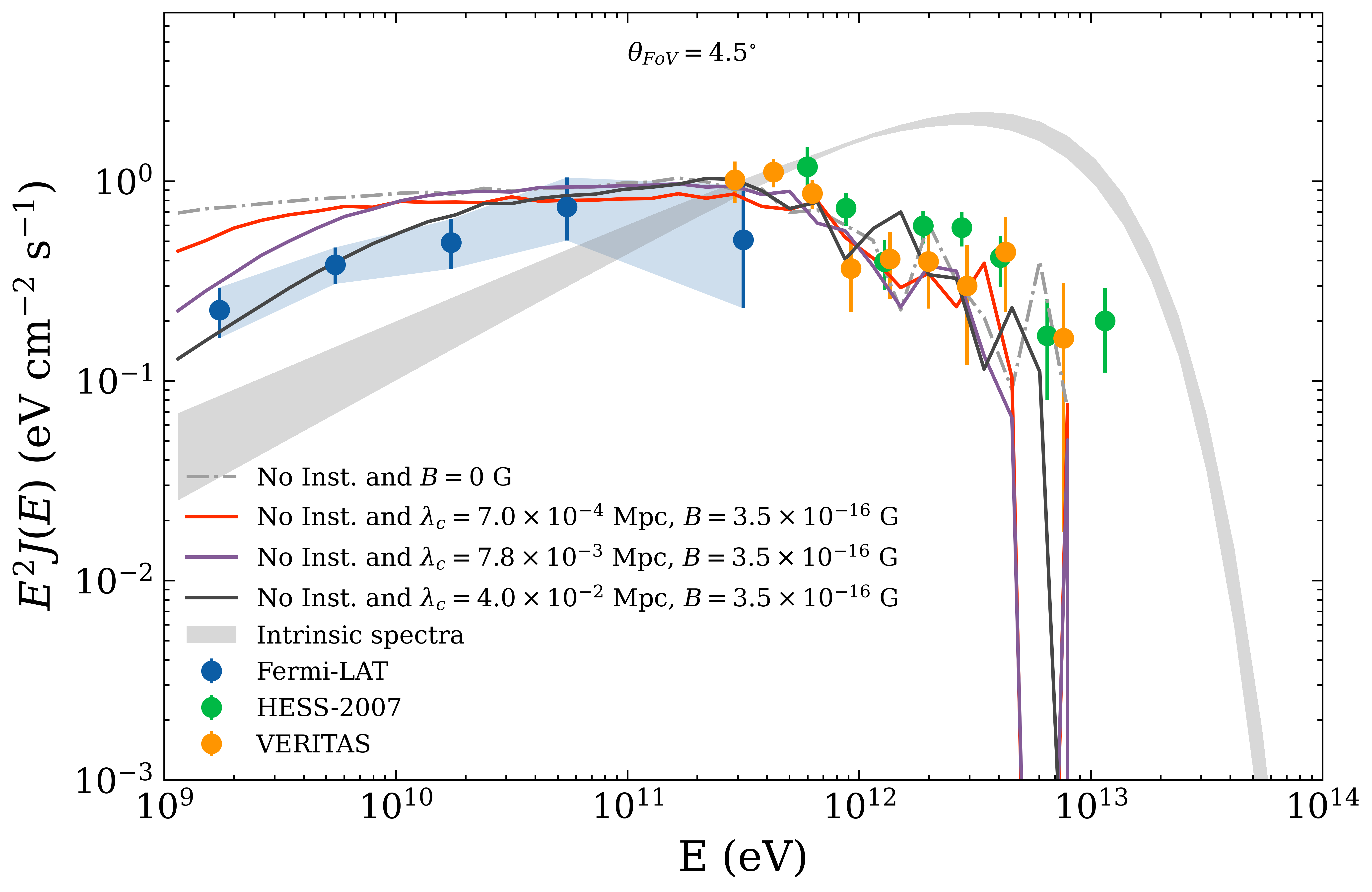}
   \includegraphics[width=0.49\textwidth]{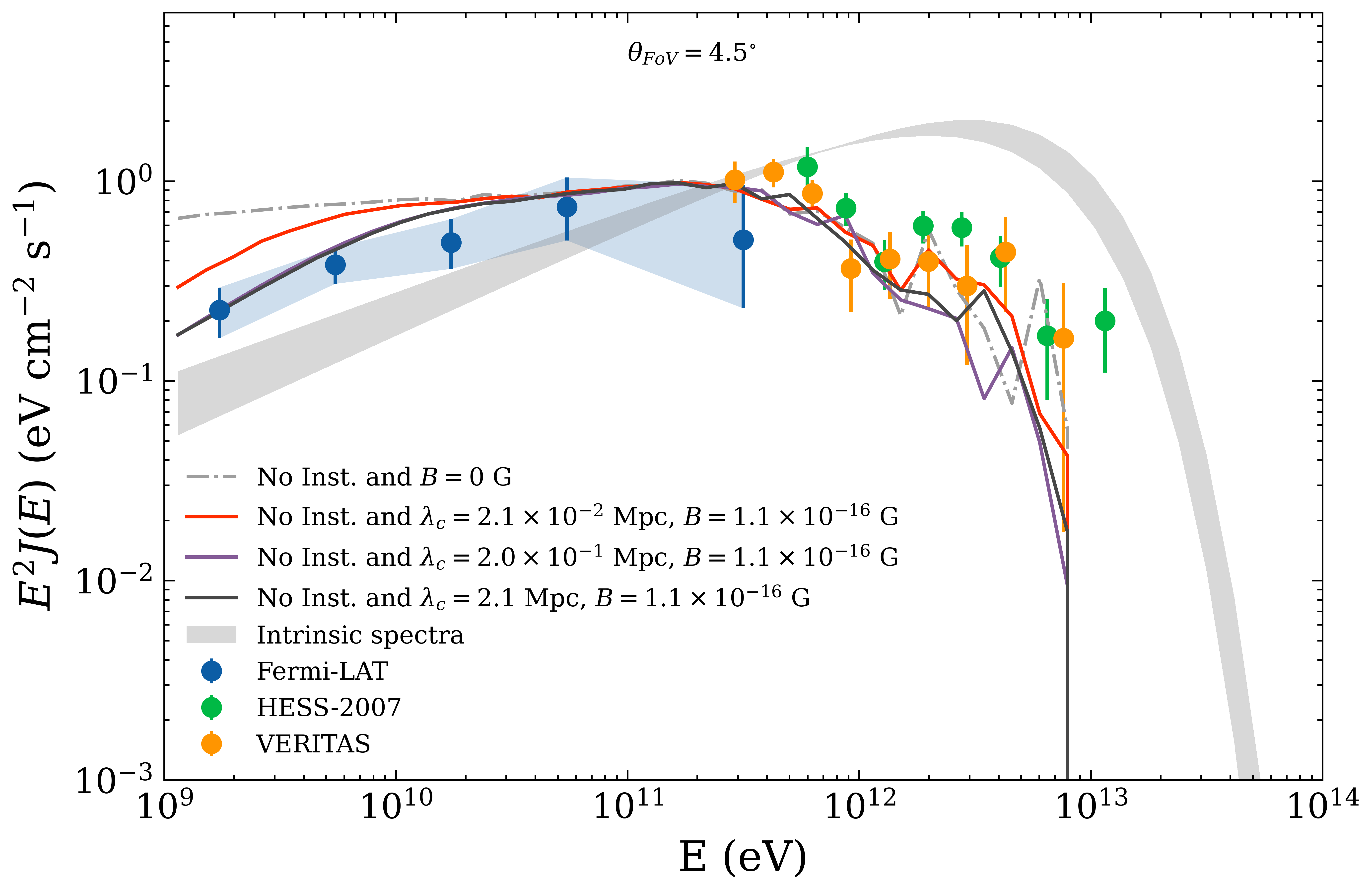}
      \caption{Same as Fig.~\ref{fig:theta-1-0-only-mag}, in this case, the spectral energy distribution is obtained for $\theta_{\text{FoV}}=4.5^{\circ}$. The corresponding injection parameters and the $\chi^{2}/n_{\text{dof}}$ test statistic for the GeV-band data from Fermi-LAT ($n_{\text{dof}}=2$) are listed in Table~\ref{tab:chi-sq2-only-mag}.}
              \label{fig:theta-4-5-only-mag}%
\end{figure*}

First,we simulate the propagated photon spectrum without taking into account the effects of plasma instabilities and magnetic deflections. In this scenario, all secondary GeV photons are generated by IC interactions of the electrons and positrons. We then perform the propagation in the presence of IGMF. The electrons and positrons undergo magnetic deflections before producing GeV photons via IC scattering. We compute the cascade signal $G(E_{0}, E; \lambda_{\text{c}}, B, \theta_{\text{FoV}})$, which is independent of the instability parameters, for each configuration of IGMF strength, $B$ and the corresponding coherence length, $\lambda_{\text{c}}$. In the first panel of Fig.~\ref{fig:theta-1-0-only-mag}, the gray dashed-dot curve represents the propagated spectrum without instability and IGMF. We then introduce IGMF with a strength of $10^{-15}$ G and vary its coherence length to determine the spectrum that best agrees with the observed data from Fermi-LAT, H.E.S.S., and VERITAS, shown as blue, orange, and green points, respectively. The colored curves in the first panel of Fig.~\ref{fig:theta-1-0-only-mag} show the propagated photon spectra for different $\lambda_{\text{c}}$ values with an IGMF strength of $10^{-15}$ G. In the second and third panels of Fig.~\ref{fig:theta-1-0-only-mag}, we decrease the magnetic field strength to values lower than in the previous case, and the coherence lengths are chosen accordingly to reproduce spectra in agreement with the observations. The parameter space is explored as follows: for the first panel, 
$B \in [0.5,15.5]\times10^{-15}\text{~G}$ with a step size 
$\Delta B = 0.1\times10^{-15}\text{~G}$, and 
$\lambda_{\text{c}} \in [10^{-4},10^{-2}]\text{~Mpc}$ with a step size 
$\Delta \log_{10}\lambda_{\text{c}} = 2 \times 10^{-3}\text{~Mpc}$; 
for the second panel, 
$B \in [0.5,15.5]\times10^{-16}\text{~G}$ with a step size 
$\Delta B = 0.1\times10^{-16}\text{~G}$, and 
$\lambda_{\text{c}} \in [10^{-4},10^{-1}]\text{~Mpc}$ with a step size 
$\Delta \log_{10}\lambda_{\text{c}} = 3 \times 10^{-3}\text{~Mpc}$; 
and for the third panel, 
$B \in [0.5,15.5]\times10^{-17}\text{~G}$ with a step size 
$\Delta B = 0.1\times10^{-17}\text{~G}$, and 
$\lambda_{\text{c}} \in [10^{-2},10^{1}]\text{~Mpc}$ with a step size 
$\Delta \log_{10}\lambda_{\text{c}} = 3 \times 10^{-3}\text{~Mpc}$, for both cases $\theta_{\text{FoV}} = 1.0^\circ$ and $4.5^\circ$. The best-fit intrinsic parameters $A,\beta, E_{\text{cut}}$ are obtained from the $\chi^{2}$ analysis using Eq.~\eqref{eq:chisq} for each $B$ and $\lambda_{\text{c}}$ combination at a 90\% confidence level, as explained in Section \ref{subsec:test-stat}. The results shown in Fig.~\ref{fig:theta-1-0-only-mag} are obtained for an observer field of view of $\theta_{\text{FoV}} = 1.0^{\circ}$. We repeat the analysis for a larger field of view angle of $\theta_{\text{FoV}} = 4.5^{\circ}$, as shown in Fig.~\ref{fig:theta-4-5-only-mag}. As we increase the field of view, the observer captures a larger angular extent of the emission by including particles with slightly greater deflection angles, and we need a slightly stronger magnetic field to reconstruct the observed spectrum.

In the first panels of Figs.~\ref{fig:theta-1-0-only-mag} and \ref{fig:theta-4-5-only-mag}, the best-fit spectrum corresponds to a larger $\lambda_{\text{c}}$, with the same IGMF strength of $10^{-15}$ G. This is because the deflection angle is proportional to the product of the magnetic field strength and the square root of coherence length, since the size of the extended emission slightly increases, which implies larger deflection angles, a slightly larger coherence length is therefore required. In the second and third panels of the same figures, we explore different combinations of IGMF strength and $\lambda_{\text{c}}$ to determine the configuration that best reproduces the observed spectrum. In particular, the third panels of Figs.~\ref{fig:theta-1-0-only-mag} and \ref{fig:theta-4-5-only-mag} show that the cascade spectra are the same for $\lambda_{\text{c}}\geq 0.2$ Mpc at a specific magnetic field strength. In this scenario, the propagation is almost ballistic, and the deflection angle is proportional to B but independent of $\lambda_{\text{c}}$. The corresponding combination of IGMF strength and coherence length thus represents the lower limit of the IGMF. Further discussion on this is provided in Section~\ref{sec:Lower-lim-with-inst}. The cumulative minimum $\chi^{2}_{\text{min}}(\lambda_{\text{c}};B)/n_{\text{dof}}$ for each propagated spectrum corresponding to a combination of $\lambda_{\text{c}}$ and $B$ are listed in Table~\ref{tab:chi-sq2-only-mag}.\\

\begin{table*}[!ht]\footnotesize\centering
\caption{$\chi^{2}_\text{min}(A, \beta, E_{\text{cut}};\lambda_{\text{c}}, B)/n_{\text{dof}}$ values are obtained for the GeV-band data from Fermi-LAT ($n_{\text{dof}}=2$), for $\theta_{\text{FoV}}=1.0^{\circ}$ and $4.5^{\circ}$. The $\chi^2_{\min}/n_{\text{dof}}
= \min_{\lambda_{\text{c}}} [ \chi^2_{\min}(A, \beta, E_{\text{cut}};\lambda_{\text{c}}, B) ]/n_{\text{dof}}$ value is highlighted in bold.}\label{tab:chi-sq2-only-mag}
\resizebox{0.85\textwidth}{!}{
\begin{tabular}{c*{8}{c}r}
 \hline
 \multirow{2}{*}{$\lambda_{0}$ [kpc]} & \multirow{2}{*}{$\theta_{\text{FoV}}$ [deg]} & \multirow{2}{*}{$B$ [G]} & \multirow{2}{*}{$\lambda_{\text{c}}$ [Mpc]} & \multirow{2}{*}{$\beta$} & \multirow{2}{*}{$E_{\text{cut}}$ [TeV]} & $\text{log}_{10}A$ & \multirow{2}{*}{$\chi_{\text{min}}^{2}/n_{\text{dof}}$}\\ & & & & & & \makecell[c]{[eV$^{-1}$cm$^{-2}$sec$^{-1}$]} \\ \hline
{} & {} & {} & $2.0\times 10^{-4}$ & $1.39$ & $6.3$ & $-23.717$ & $3.311$ \\
{} & {} & $10^{-15}$ & $\mathbf{9.0\times 10^{-4}}$ & $\mathbf{1.48}$ & $\mathbf{5.3}$ & $\mathbf{-23.708}$ & $\mathbf{2.350}$ \\
{} & {} & {} & $7.0\times 10^{-3}$ & $1.54$ & $6.4$ & $-23.711$ & $3.040$ \\ \cline{3-8}
{} & {} & {} & $8.0\times 10^{-4}$ & $1.46$ & $5.3$ & $-23.708$ & $3.266$ \\
{} & $1.0$ & $3.0\times 10^{-16}$ & $\mathbf{9.8\times 10^{-3}}$ & $\mathbf{1.57}$ & $\mathbf{4.6}$ & $\mathbf{-23.718}$ & $\mathbf{2.287}$ \\
{} & {} & {} & $6.8\times 10^{-2}$ & $1.41$ & $5.8$ & $-23.712$ & $2.606$ \\ \cline{3-8}
{} & {} & {} & $2.0\times 10^{-2}$ & $1.43$ & $5.0$ & $-23.722$ & $2.331$ \\
{} & {} & $7.0\times 10^{-17}$ & $\mathbf{2.0\times 10^{-1}}$ & $\mathbf{1.51}$ & $\mathbf{5.5}$ & $\mathbf{-23.719}$ & $\mathbf{2.188}$ \\
\multirow{1}{*}{No} & {} & {} & $\mathbf{2.1}$ & $\mathbf{1.53}$ & $\mathbf{5.2}$ & $\mathbf{-23.711}$ & $\mathbf{2.188}$ \\ \cline{2-8}
 
\multirow{1}{*}{Inst.} & {} & {} & $8.7\times 10^{-4}$ & $1.40$ & $5.5$ & $-23.718$ & $2.707$ \\
{} & {} & $10^{-15}$ & $\mathbf{4.0\times 10^{-3}}$ & $\mathbf{1.45}$ & $\mathbf{5.0}$ & $\mathbf{-23.726}$ & $\mathbf{1.919}$ \\
{} & {} & {} & $6.0\times 10^{-2}$ & $1.50$ & $5.2$ & $-23.714$ & $2.135$ \\ \cline{3-8}
{} & {} & {} & $7.0\times 10^{-4}$ & $1.36$ & $5.1$ & $-23.717$ & $3.508$ \\
{} & $4.5$ & $3.5\times 10^{-16}$ & $7.8\times 10^{-3}$ & $1.42$ & $5.9$ & $-23.711$ & $2.972$ \\
{} & {} & {} & $\mathbf{4.0\times 10^{-2}}$ & $\mathbf{1.51}$ & $\mathbf{5.7}$ & $\mathbf{-23.723}$ & $\mathbf{2.833}$ \\ \cline{3-8}
{} & {} & {} & $2.1\times 10^{-2}$ & $1.47$ & $5.6$ & $-23.714$ & $3.054$ \\
{} & {} & $1.1\times 10^{-16}$ & $\mathbf{2.0\times 10^{-1}}$ & $\mathbf{1.56}$ & $\mathbf{4.9}$ & $\mathbf{-23.720}$ & $\mathbf{2.549}$ \\
{} & {} & {} & $\mathbf{2.1}$ & $\mathbf{1.58}$ & $\mathbf{4.8}$ & $\mathbf{-23.718}$ & $\mathbf{2.549}$ \\ \hline
\end{tabular}}
\end{table*}

\subsection{Modified electromagnetic cascade with IGMF and plasma instability}
\begin{figure*}[!ht]
   \centering
   \includegraphics[width=0.49\textwidth]{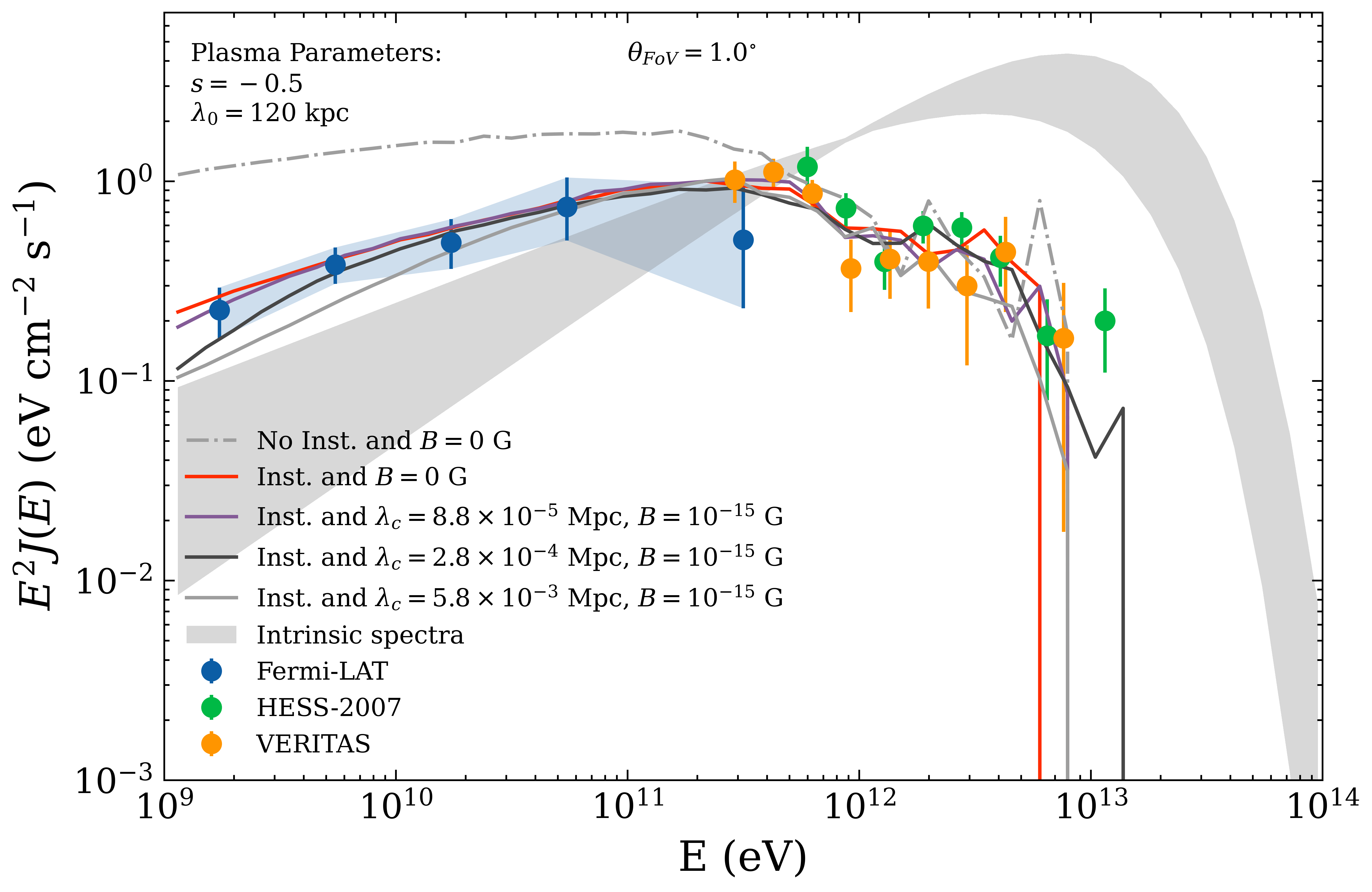}
   \includegraphics[width=0.49\textwidth]{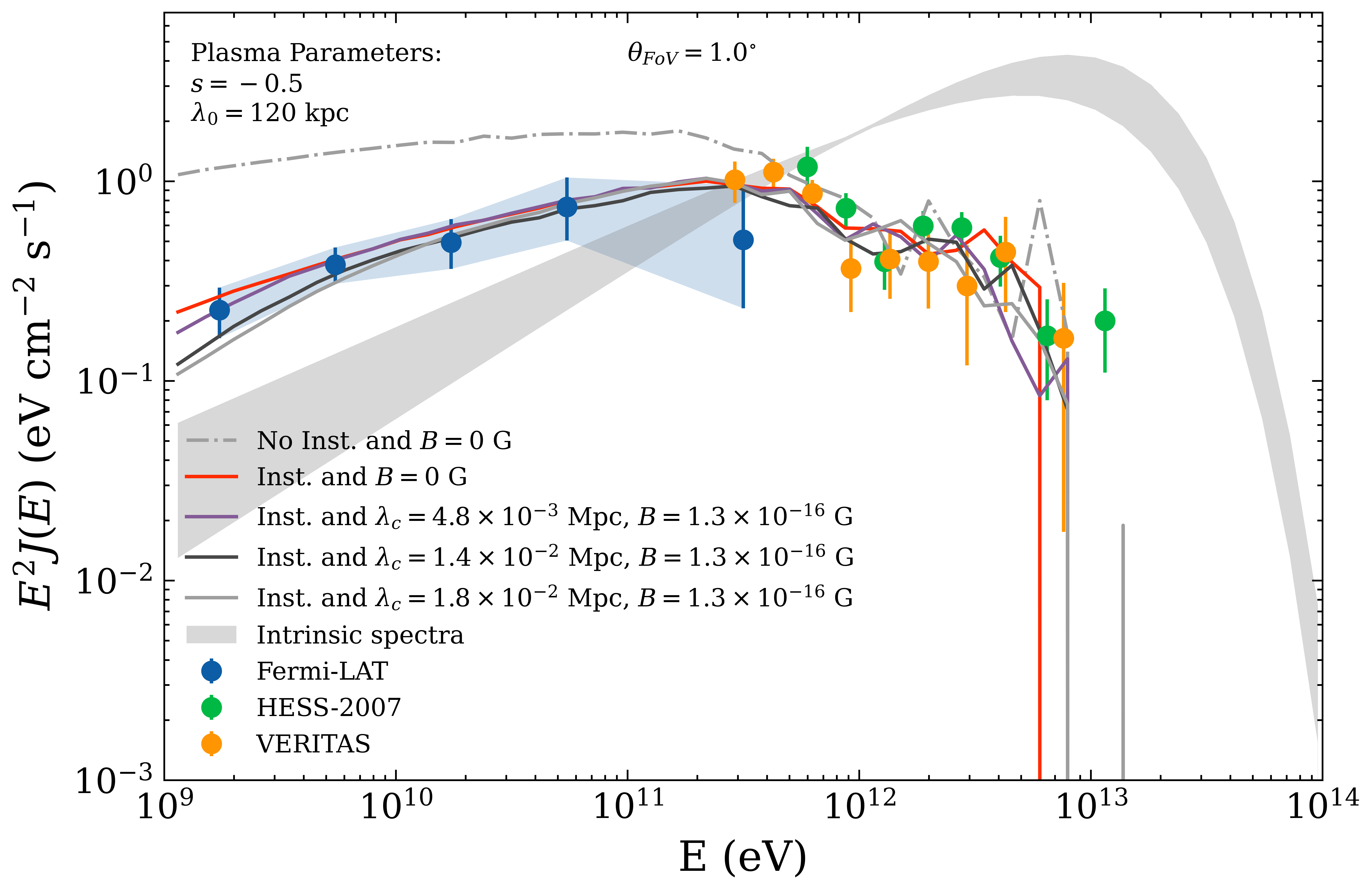}
   \includegraphics[width=0.49\textwidth]{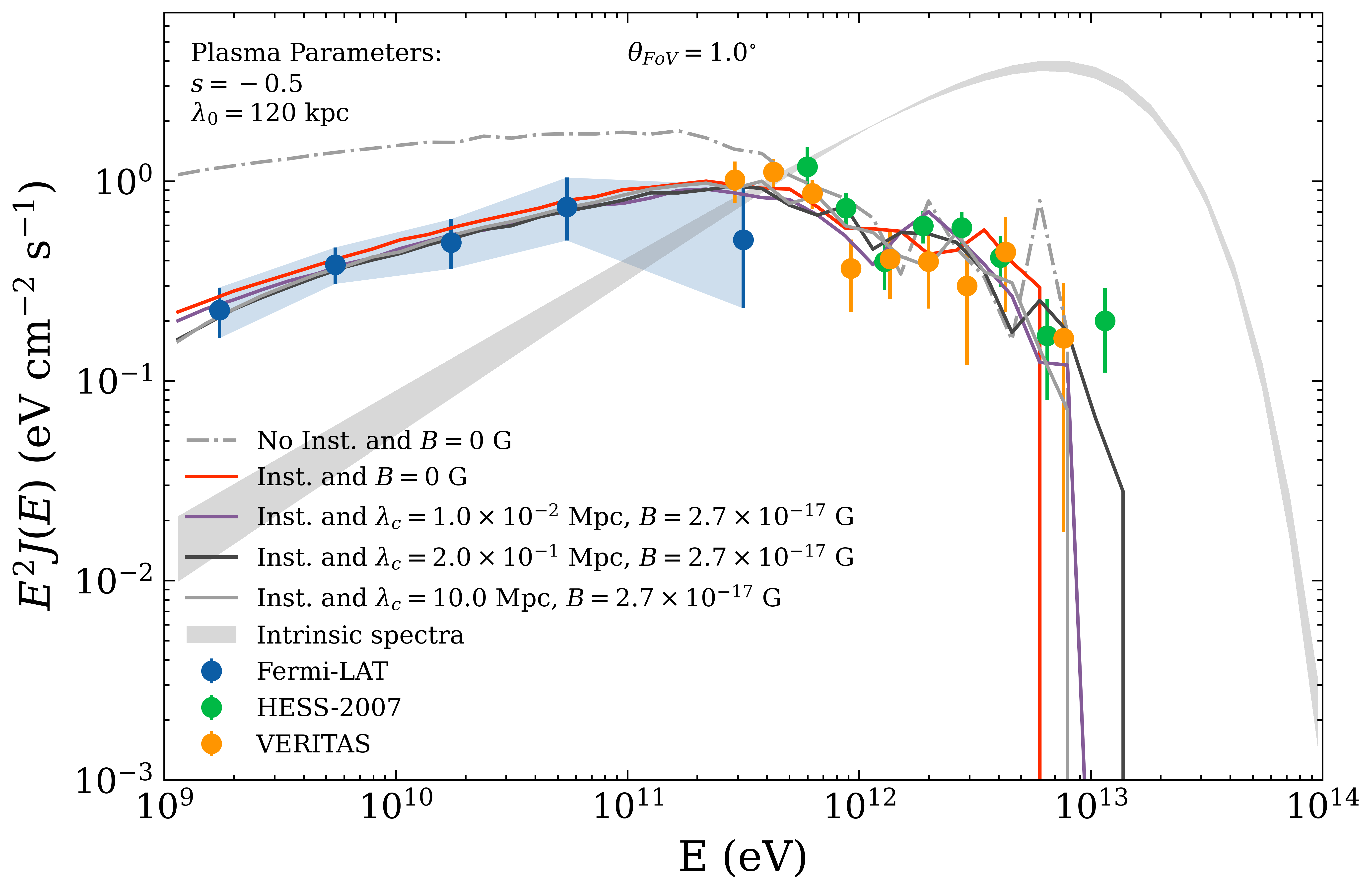}
   \caption{The energy spectrum of 1ES 0229+200 in the energy range $10^{-3}\leq E/\text{TeV} \leq 10^{2}$. The gray dash-dotted curve shows the propagated photon spectrum without any contribution of the plasma instability cooling (Inst. stands for Instability in the plot legends) and IGMF. The colored solid curves represent the propagated photon spectrum for different IGMF strengths and coherence length combinations, as well as plasma instability cooling for $\lambda_{0} = 120\text{~kpc}$ and $\alpha = -0.5$. The plasma-instability parameters are adopted from the best-fit scenario obtained in the previous subsection (also consistent with the results of \cite{Castro:2024ooo}). The gray band shows the unattenuated intrinsic spectra, while the corresponding injection parameters are listed in Table~\ref{tab:chi-sq2-lam0-120}. The blue data points indicate the Fermi-LAT \cite{MAGIC:2022piy}, the green data points show the H.E.S.S. \cite{HESS:2007xak}, and the orange data points show the VERITAS \cite{Aliu:2013pya} spectrum. We investigate the energy spectrum, which is consistent with the Fermi-LAT, H.E.S.S., and VERITAS observational data in each plot. The energy spectrum is obtained for a fixed $\theta_{\text{FoV}}=1.0^{\circ}$. The $\chi^{2}/n_{\text{dof}}$ test statistic for the GeV-band data from Fermi-LAT ($n_{\text{dof}}=2$) is summarized in Table~\ref{tab:chi-sq2-lam0-120}.}
              \label{fig:theta-1-0-lam0-120}%
\end{figure*}

\begin{figure*}[!ht]
   \centering
   \includegraphics[width=0.49\textwidth]{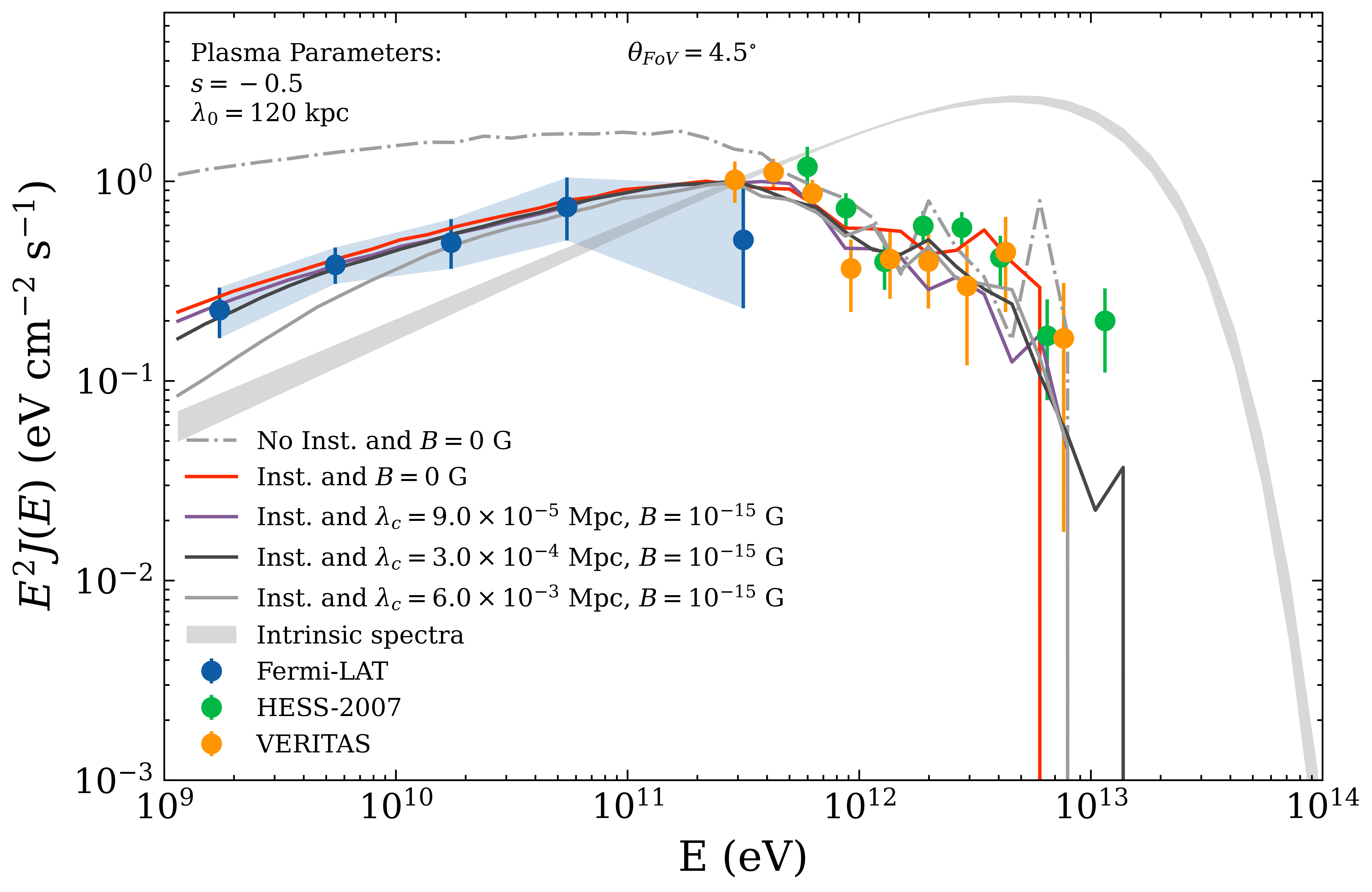}
   \includegraphics[width=0.49\textwidth]{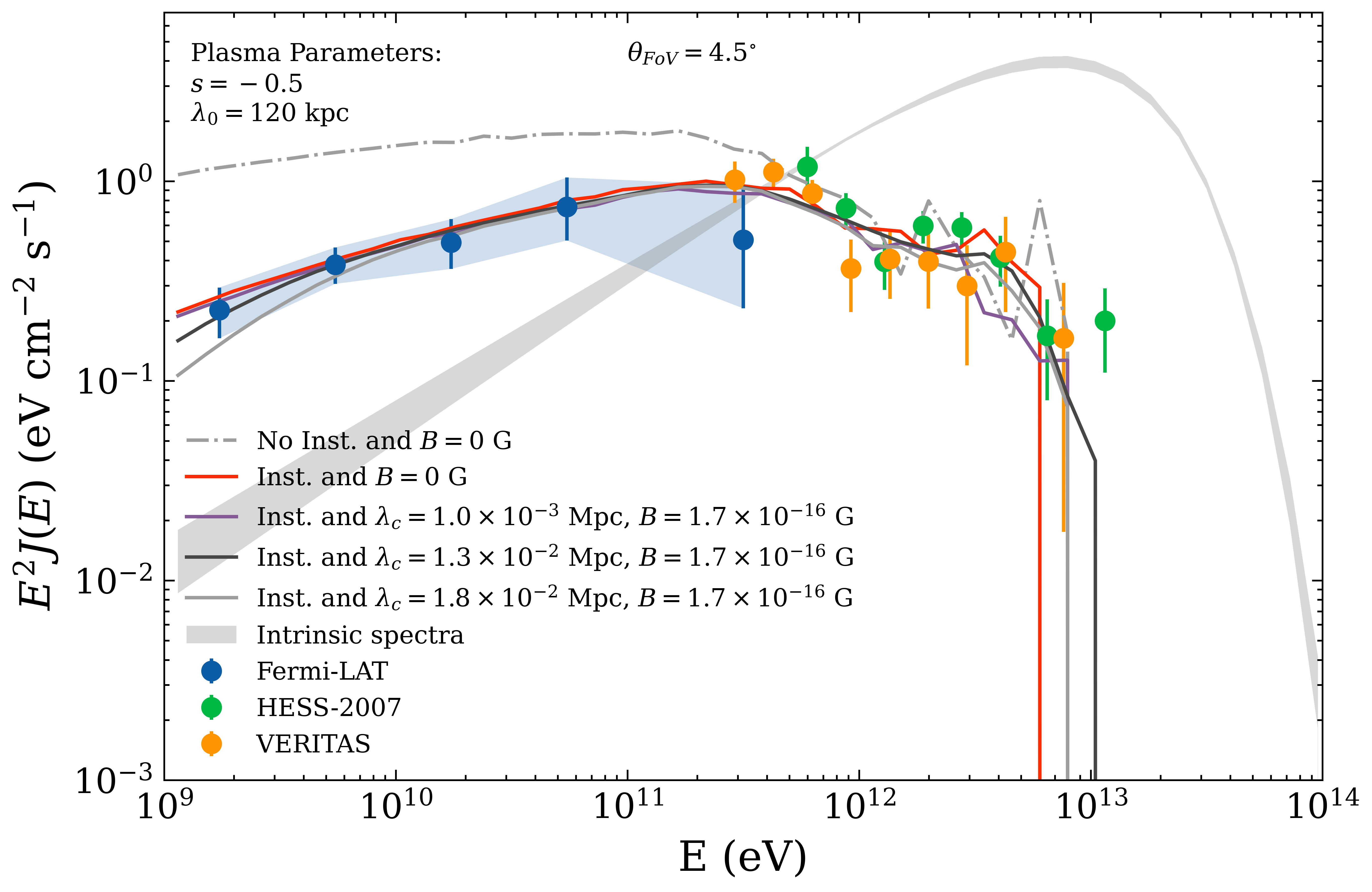}
   \includegraphics[width=0.49\textwidth]{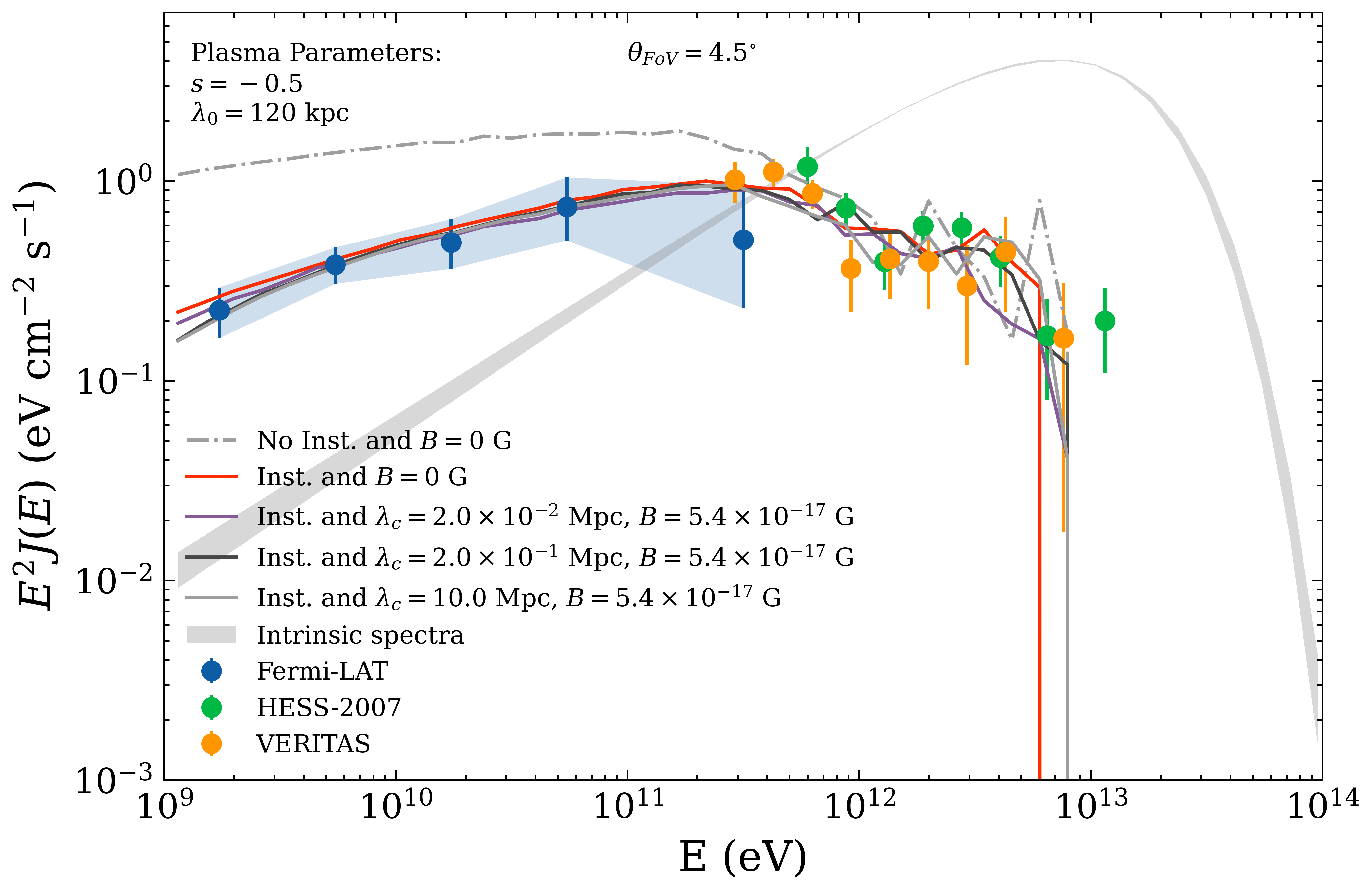}
      \caption{Same as Fig.~\ref{fig:theta-1-0-lam0-120}, in this case, the spectral energy distribution is obtained for $\theta_{\text{FoV}}=4.5^{\circ}$. The corresponding injection parameters and the $\chi^{2}/n_{\text{dof}}$ test statistic for the GeV-band data from Fermi-LAT ($n_{\text{dof}}=2$) are listed in Table~\ref{tab:chi-sq2-lam0-120}.}
              \label{fig:theta-4-5-lam0-120}%
\end{figure*}

In this section, we investigate the combined effect of plasma instability-induced energy losses and the propagation of electrons (and positrons) in the presence of IGMF, to reproduce the observed photon spectrum of the blazar 1ES 0229+200. We obtain the propagated photon spectrum from the cascade signal $G(E_{0}, E; \lambda_{0}, \alpha; \lambda_{\text{c}}, B, \theta_{\text{FoV}})$, as defined in Eq.~\ref{eq:reconstruction-spectrum}, for each configuration of IGMF strength $B$, and corresponding coherence length $\lambda_{\text{c}}$, including the instability parameters $\lambda_{0}$ and $\alpha$. e consider the best-fit instability parameters obtained in \ref{appen:cascade-with-instability-only}, $\lambda_{0}=120\text{~kpc}$ and $\alpha=-0.5$, as the key parameter determining whether the cascade is dominated by plasma instabilities or by IC scattering.
Following the approach of the previous section, we start from the propagated spectrum obtained without IGMF and instability energy losses, represented by the gray dashed curve in Fig.~\ref{fig:theta-1-0-lam0-120}. In the same figure, the red solid curve represents the spectrum including instability cooling on a typical length scale of $\lambda_{0} = 120$ kpc, without any contribution from the IGMF. We see that the plasma instabilities already suppress the propagated photon spectrum in the GeV energy range in the absence of an IGMF (as shown by the red curve in the following figure). In the first panel of Fig. \ref{fig:theta-1-0-lam0-120}, we again compare the propagated photon spectra, following the same procedure as in the previous section, for different choices of $\lambda_{\text{c}}$ for $B = 10^{-15}$ G (indicated by the colored curves), this time including the instability cooling. This allows us to investigate how the combined influence of magnetic deflections and instability cooling shapes the spectrum and to identify the configuration that best reproduces the observational data. We explore the parameter space as follows: for the first panel, 
$B \in [0.5,15.5]\times10^{-15}\text{~G}$ with a step size 
$\Delta B = 0.1\times10^{-15}\text{~G}$, and 
$\lambda_{\text{c}} \in [10^{-5},10^{-3}]\text{~Mpc}$ with a step size 
$\Delta \log_{10}\lambda_{\text{c}} = 2 \times 10^{-3}\text{~Mpc}$; 
for the second panel, 
$B \in [0.5,15.5]\times10^{-16}\text{~G}$ with a step size 
$\Delta B = 0.1\times10^{-16}\text{~G}$, and 
$\lambda_{\text{c}} \in [10^{-3},10^{-2}]\text{~Mpc}$ with a step size 
$\Delta \log_{10}\lambda_{\text{c}} =10^{-3}\text{~Mpc}$; 
and for the third panel, 
$B \in [0.5,15.5]\times10^{-17}\text{~G}$ with a step size 
$\Delta B = 0.1\times10^{-17}\text{~G}$, and 
$\lambda_{\text{c}} \in [10^{-2},10^{1}]\text{~Mpc}$ with a step size 
$\Delta \log_{10}\lambda_{\text{c}} = 3 \times 10^{-3}\text{~Mpc}$, for both cases $\theta_{\text{FoV}} = 1.0^\circ$ and $4.5^\circ$. In the same way as in the previous section, the second and third panels of Fig.~\ref{fig:theta-1-0-lam0-120} show results obtained by reducing the magnetic field strength and choosing the corresponding coherence lengths such that the resulting spectra remain consistent with the observations. The best-fit prompt emission parameters $A,\beta, E_{\text{cut}}$ are obtained from the $\chi^{2}$ analysis using Eq.~\eqref{eq:chisq} for each ($B, \lambda_{\text{c}}$) combination at the 90\% confidence level, as described in Section~\ref{subsec:test-stat}. We obtain the photon spectrum shown in Fig.~\ref{fig:theta-1-0-lam0-120} for an observer field of view of $\theta_{\text{FoV}} = 1.0^{\circ}$. We then repeat the analysis for a larger field of view, $\theta_{\text{FoV}} = 4.5^{\circ}$, following the same procedure and for the same instability cooling scale (see Fig.~\ref{fig:theta-4-5-lam0-120}). The cumulative minimum $\chi^{2}_{\text{min}}(\lambda_{\text{c}};B)/n_{\text{dof}}$ for each propagated spectrum corresponding to a combination of $\lambda_{\text{c}}$ and $B$ are listed in Table \ref{tab:chi-sq2-lam0-120}.

\begin{table*}[!ht]\footnotesize\centering
\caption{$\chi^{2}_\text{min}(A, \beta, E_{\text{cut}};\lambda_{\text{c}}, B)/n_{\text{dof}}$ values  are obtained for the GeV-band data from Fermi-LAT ($n_{\text{dof}}=2$), for $\theta_{\text{FoV}}=1.0^{\circ}$ and $4.5^{\circ}$ and best-fit plasma instability parameters, $\lambda_{0}=120$ kpc, $\alpha=-0.5$. The $\chi^2_{\min}/n_{\text{dof}}
= \min_{\lambda_{\text{c}}} [ \chi^2_{\min}(A, \beta, E_{\text{cut}};\lambda_{\text{c}}, B) ]/n_{\text{dof}}$ value are highlighted in bold.}\label{tab:chi-sq2-lam0-120}
\resizebox{0.95\textwidth}{!}{
\begin{tabular}{c*{9}{c}r}
 \hline
 \multirow{2}{*}{$\lambda_{0}$ [kpc]} & \multirow{2}{*}{$\alpha$} & \multirow{2}{*}{$\theta_{\text{FoV}}$ [deg]} & \multirow{2}{*}{$B$ [G]} & \multirow{2}{*}{$\lambda_{\text{c}}$ [Mpc]} & \multirow{2}{*}{$\beta$} & \multirow{2}{*}{$E_{\text{cut}}$ [TeV]} & $\text{log}_{10}A$ & \multirow{2}{*}{$\chi_{\text{min}}^{2}/n_{\text{dof}}$}\\ & & & & & & & \makecell[c]{[eV$^{-1}$cm$^{-2}$sec$^{-1}$]} \\ \hline
{} & {} & {} & {} & $\mathbf{8.8\times 10^{-5}}$ & $\mathbf{1.27}$ & $\mathbf{10.5}$ & $\mathbf{-23.705}$ & $\mathbf{1.613}$ \\
{} & {} & {} & $10^{-15}$ & $2.8\times 10^{-4}$ & $1.20$ & $9.5$ & $-23.720$ & $1.688$ \\
{} & {} & {} & {} & $5.8\times 10^{-3}$ & $1.55$ & $7.7$ & $-23.710$ & $1.918$ \\ \cline{4-9}
{} & {} & {} & {} & $\mathbf{4.8\times 10^{-3}}$ & $\mathbf{1.28}$ & $\mathbf{10.7}$ & $\mathbf{-23.711}$ & $\mathbf{1.577}$ \\
{} & {} & $1.0$ & $1.3\times 10^{-16}$ & $1.4\times 10^{-2}$ & $1.31$ & $9.9$ & $-23.719$ & $1.680$ \\
{} & {} & {} & {} & $1.8\times 10^{-2}$ & $1.49$ & $10.1$ & $-23.712$ & $1.750$ \\ \cline{4-9}
{} & {} & {} & {} & $1.0\times 10^{-2}$ & $1.22$ & $8.9$ & $-23.713$ & $1.194$ \\
{} & {} & {} & $2.7\times 10^{-17}$ & $\mathbf{2.0\times 10^{-1}}$ & $\mathbf{1.31}$ & $\mathbf{9.5}$ & $\mathbf{-23.711}$ & $\mathbf{1.191}$ \\
\multirow{2}{*}{$120$} & \multirow{2}{*}{$-0.5$} & {} & {} & $\mathbf{10.0}$ & $\mathbf{1.33}$ & $\mathbf{10.1}$ & $\mathbf{-23.710}$ & $\mathbf{1.191}$ \\ \cline{3-9}
 
{} & {} & {} & {} & $9.0\times 10^{-5}$ & $1.51$ & $9.2$ & $-23.713$ & $1.310$ \\
{} & {} & {} & $10^{-15}$ & $\mathbf{3.0\times 10^{-4}}$ & $\mathbf{1.48}$ & $\mathbf{9.7}$ & $\mathbf{-23.710}$ & $\mathbf{1.306}$ \\
{} & {} & {} & {} & $6.0\times 10^{-3}$ & $1.46$ & $8.7$ & $-23.717$ & $1.606$ \\ \cline{4-9}
{} & {} & {} & {} & $1.0\times 10^{-3}$ & $1.20$ & $9.0$ & $-23.711$ & $1.357$ \\
{} & {} & $4.5$ & $1.7\times 10^{-16}$ & $\mathbf{1.3\times 10^{-2}}$ & $\mathbf{1.24}$ & $\mathbf{9.6}$ & $\mathbf{-23.702}$ & $\mathbf{1.352}$ \\
{} & {} & {} & {} & $1.8\times 10^{-2}$ & $1.31$ & $10.3$ & $-23.719$ & $1.449$ \\ \cline{4-9}
{} & {} & {} & {} & $2.0\times 10^{-2}$ & $1.21$ & $8.9$ & $-23.715$ & $1.301$ \\
{} & {} & {} & $5.4\times 10^{-17}$ & $\mathbf{2.0\times 10^{-1}}$ & $\mathbf{1.27}$ & $\mathbf{10.1}$ & $\mathbf{-23.713}$ & $\mathbf{1.297}$ \\
{} & {} & {} & {} & $\mathbf{10.0}$ & $\mathbf{1.25}$ & $\mathbf{9.7}$ & $\mathbf{-23.711}$ & $\mathbf{1.297}$ \\ \hline
\end{tabular}}
\end{table*}
We find that the plasma instability scenario for $\lambda_{0} \geq 120$ kpc (with $\alpha = -0.5$) represents the threshold instability loss length required to reconstruct the photon spectrum in agreement with observations. For $\lambda_{0}$ below this value, even in the absence of an IGMF, the propagated spectra fall below the observational data in the GeV energy range. We evaluate the fractional energy loss due to instability over a distance of a single IC interaction length in the presence of the IGMF. If $E_{e}(x)$ is the energy of an electron at a distance $x$ from the point of injection, then the energy loss of the electron within an IC interaction length, $\lambda_{\text{IC}}$, is given by $\Delta E_{e}(\lambda_{\text{IC}}) = E_{e}(x_0) - E_{e}(\lambda_{\text{IC}})$, where $x_0$ is the injection point. Then, the fractional energy loss due to instability over a distance of a single IC interaction length is given by \cite{Castro:2024ooo}
\begin{equation}
    \left(\frac{\Delta E_{e}(\lambda_{\text{IC}})}{E_{e}(x_0)}\right)_{\text{Inst.}} = \frac{\lambda_{\text{IC}}}{\lambda_{0}}\cdot\left(\frac{E_{e}(x_0)}{1\text{~TeV}}\right)^{1/2}.
    \label{eq:frac-loss}
\end{equation}
In the plasma instability scenario with $\lambda_{0} = 120$ kpc (for $\alpha = -0.5$), the fractional energy loss within a single IC interaction length remains below $2$\% for electrons with energies below $4$ TeV, and less than about $1$\% for electrons in the GeV energy range. Overall, the effect of the instability is marginal, even in the best-fit scenario.

\section{Impact of instability on IGMF constraint}\label{sec:Lower-lim-with-inst}
Relativistic electrons with energy $0.6\lesssim E_{e}/\text{TeV} \lesssim 5.6$ are produced by the TeV photons that interact with CMB photons \cite{Neronov:2009gh}. These electrons lose energy through IC scattering over a characteristic cooling length scale of $0.1\lesssim l_{\text{IC}}/\text{Mpc}\lesssim 0.6$ \cite{Neronov:2009gh}. If the magnetic coherence length is longer than the length scale of electron cooling via IC ($l_{\text{IC}}<\lambda_{\text{c}}$), then the deflection angle of the $e^{+}e^{-}$ pairs, $\delta = l_{\text{IC}}/r_{L}$, becomes independent of the coherence length. Here, $r_{L} = E_{e}/eB$ is the Larmor radius in a magnetic field $B$. In contrast, in the regime $l_{\text{IC}}>\lambda_{\text{c}}$, the deflection follows a random walk, and particles experience random magnetic field orientations, causing a diffusive angular spread. In this case, the field geometry becomes turbulent (many small cells), causing random-walk deflection with a deflection angle, $\delta_{\text{cell}} \simeq \lambda_{\text{c}}/r_{L}$. After $n=l_{\text{IC}}/\lambda_{\text{c}}$ cells, the cumulative effect of these small deflections over $l_{\text{IC}}$ distance leads to an overall deflection angle, $\delta \simeq \delta_{\text{cell}}\sqrt{l_{\text{IC}}/\lambda_{\text{c}}} = eB\sqrt{l_{\text{IC}}\lambda_{\text{c}}}/E_{e}\propto B\sqrt{\lambda_{\text{c}}}$ \cite{Vovk:2023qfk, Korochkin:2020pvg}. Thus, the lower bound can be interpreted when the field is independent of $\lambda_{\text{c}}$ when $\lambda_{\text{c}}> l_{\text{IC}}$, while for shorter coherence lengths ($\lambda_{\text{c}}< l_{\text{IC}}$) the field strength scales as $\lambda_{\text{c}}^{-0.5}$.
\begin{figure*}[!ht]
   \centering
   \includegraphics[width=0.45\textwidth]{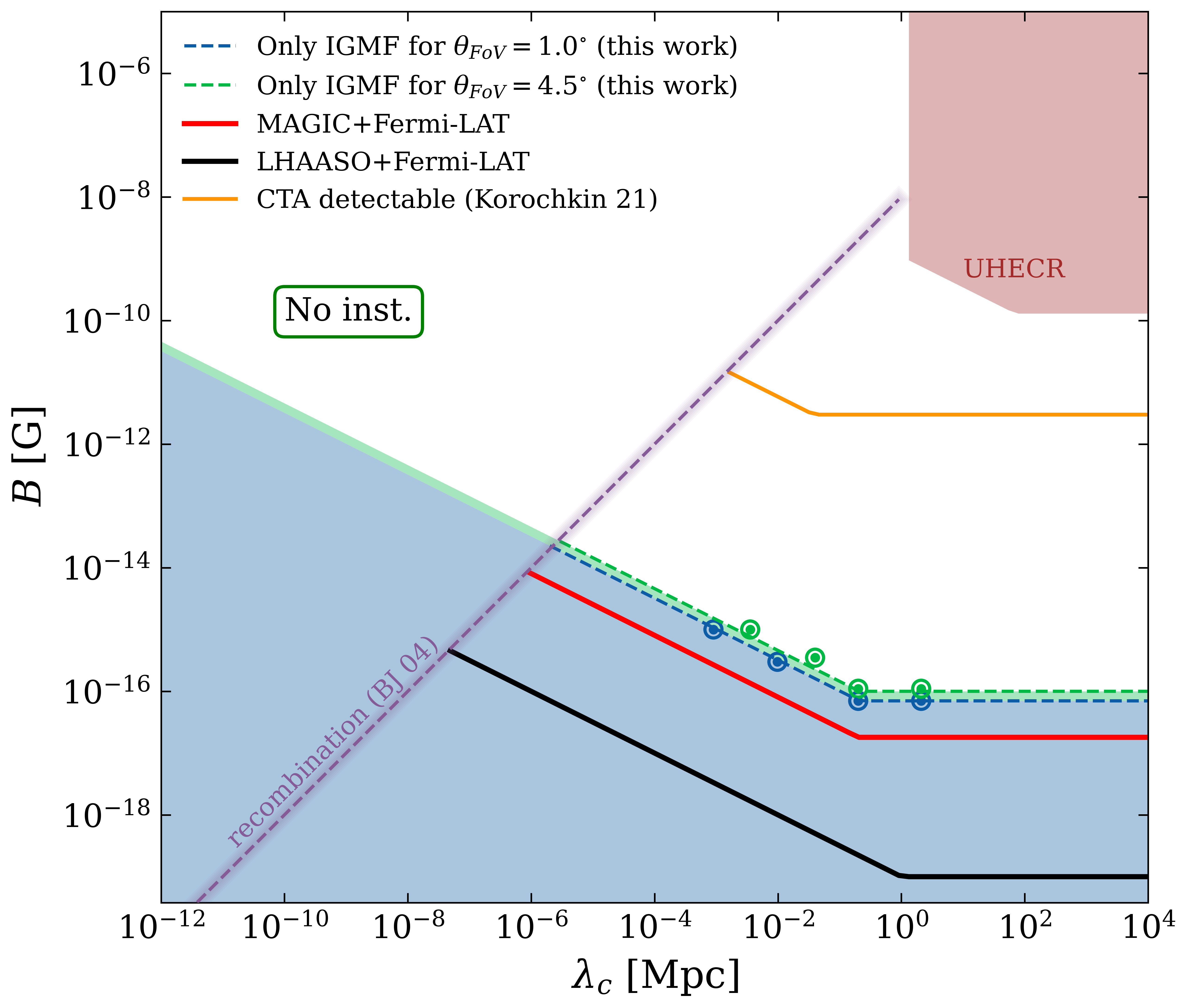}
   \includegraphics[width=0.45\textwidth]{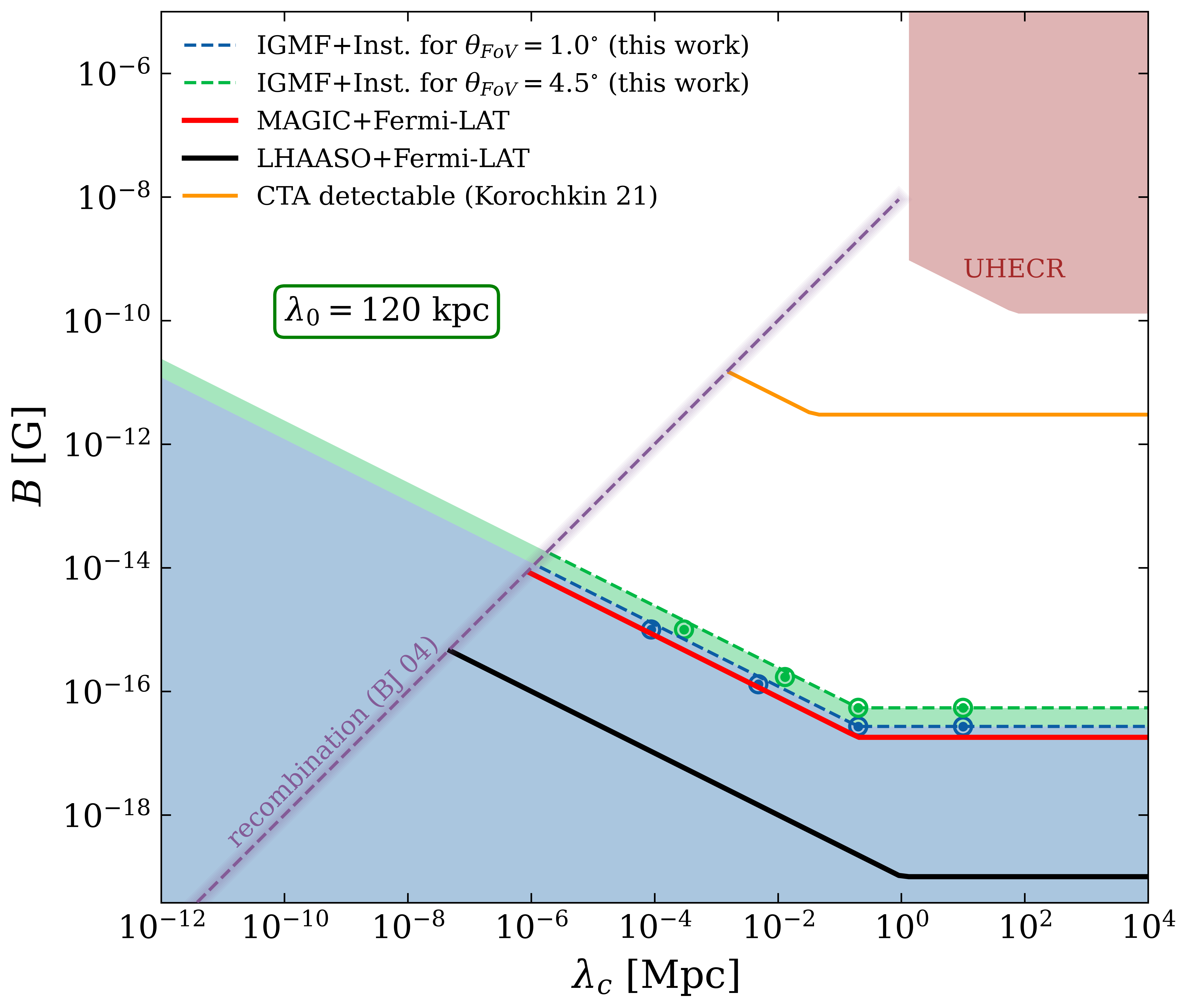}
      \caption{The lower limits on the IGMF inferred from blazar 1ES 0229+200, with (right panel) and without (left panel) plasma instabilities, are shown alongside existing constraints from MAGIC+Fermi-LAT \cite{MAGIC:2022piy} (red line), LHAASO+Fermi-LAT observations of GRB 221009A \cite{Vovk:2023qfk} (black line), UHECR measurements \cite{Neronov:2021xua} (light brown shaded region), predictions from cosmological primordial magnetic-field evolution models \cite{Banerjee:2004df} (purple dashed line), and IGMF limits that will be detectable by the Cherenkov Telescope Array (CTA) \cite{Korochkin:2020pvg} (orange line). The green and blue dashed curves represent the limits derived for observational fields of view $\theta_{\text{FoV}}=1.0^{\circ}$ and $4.5^{\circ}$, respectively. The left panel shows the IGMF constraints obtained without including plasma instabilities. In contrast, the right panel shows the modified constraints on IGMF for instability length scales approximately two orders of magnitude larger than the IC scattering length, respectively, with $\lambda_{0} = 120$ kpc and an instability spectral index of $\alpha = -0.5$.}
              \label{B-lambda-c}%
\end{figure*}
\begin{table*}[ht]\centering
\caption{The summary of IGMF limits with the best-fit plasma instability parameters.}\label{tab:summarized}
\resizebox{1.0\textwidth}{!}{
\begin{tabular}{c*{6}{c}r}
 \hline
{} & $\theta_{\text{FoV}}=1.0^{\circ}$ & $\theta_{\text{FoV}}=4.5^{\circ}$ & {}\\ \hline 
\makecell{\text{No Inst.}} & $B\gtrsim\begin{cases}
			7.0\times 10^{-17}\text{~G}, & \\
            7.0\times 10^{-17}(\lambda_{\text{c}}/0.20 \text{~Mpc})^{-1/2} \text{~G}, &
		 \end{cases}$ & $\begin{cases}
			1.1\times 10^{-16}\text{~G}, & \lambda_{\text{c}}> 0.20 \text{~Mpc}\\
            1.1\times 10^{-16}(\lambda_{\text{c}}/0.20 \text{~Mpc})^{-1/2} \text{~G}, & \lambda_{\text{c}}< 0.20 \text{~Mpc}
		 \end{cases}$ \\ \hline
\makecell{$\lambda_{0}=\text{120}\text{~kpc}$ and $\alpha =-0.5$} & $B\gtrsim\begin{cases}
			2.7\times 10^{-17}\text{~G}, \\
            2.7\times 10^{-17}(\lambda_{\text{c}}/0.20 \text{~Mpc})^{-1/2} \text{~G},
		 \end{cases}$ & $\begin{cases}
			5.4\times 10^{-17}\text{~G}, & \lambda_{\text{c}}> 0.20 \text{~Mpc}\\
            5.4\times 10^{-17}(\lambda_{\text{c}}/0.20 \text{~Mpc})^{-1/2} \text{~G}, & \lambda_{\text{c}}< 0.20 \text{~Mpc}
		 \end{cases}$ \\ \hline 
\end{tabular}}
\end{table*}
Fig. \ref{B-lambda-c} shows that, in the absence of plasma instabilities, the lower limit on the IGMF is $B\simeq7\times10^{-17}~\text{G}$ for an observational field of view of $\theta_{\text{FoV}}=1.0^{\circ}$, and increases to $B\simeq1.1\times10^{-16}~\text{G}$ for $\theta_{\text{FoV}}=4.5^{\circ}$. When plasma instabilities are taken into account, the lower bound of the IGMF weakens. The beam--plasma instabilities drain the energy of pair beams locally into the IGM as heat before they upscatter CMB photons to GeV energies. As a result, the observational lower bound on IGMF becomes weaker because beam--plasma instability provides an alternative, non-radiative cooling mechanism which suppresses the flux. The IGMF limits corresponding to different plasma instability scenarios are summarized in Table \ref{tab:summarized}. The red lines in Fig.~\ref{B-lambda-c} represent the lower bound of IGMF from the blazar 1ES 0229+200 observation, derived by the MAGIC+Fermi-LAT collaboration \cite{MAGIC:2022piy} using a time-delay method with a maximum geometric delay of $\tau_{\text{delay}} = 10$ years and assuming no instability-induced energy losses. The black lines in Fig. \ref{B-lambda-c} show the IGMF lower limits inferred from GRB 221009A, derived from the LHAASO+Fermi-LAT data \cite{Vovk:2023qfk} using the same time-delay method as in the previous analysis, and likewise assuming no instability-induced energy losses. As shown in the first panel of Fig.~\ref{B-lambda-c}, the angular broadening analysis provides slightly stronger constraints than the time-delay analysis. This can be understood using the analytical expressions of time delay and the angles with respect to the line of sight. Analytically, for $\lambda_{\text{c}}<l_{\text{IC}}$ and at redshifts $z\ll 1$, the delayed cascade emission exhibits a characteristic energy dependence of $\tau_{\text{delay}}\propto E^{-5/2}$, while the angular extent of the emission scales as $\theta_{\text{obs}}\propto E^{-1}$ \cite{Neronov:2009gh, Taylor:2011bn, AlvesBatista:2021sln}. As an example, for a source at $z \sim 0.14$, if the cascade spectrum is inferred from the size of the extended emission method, then for photons with energies $E\sim 10 \text{~GeV}$ within an angle with respect to the line of sight, $\theta_{\text{obs}}\sim 0.3^{\circ}$ leads to a lower limit of $B\sim 10^{-17}\text{~G}$. In contrast, if we measure the delayed cascade emission for the same energy cascade photons with a typical delay of $\tau_{\text{delay}}\sim 10\text{~years}$, it results in a weaker lower limit of $B\sim 10^{-18}\text{~G}$. This difference arises from the distinct characteristic energy scalings inherent to the two approaches. The blue and green lines in Fig.~\ref{B-lambda-c} show the IGMF lower limits with and without plasma instabilities, respectively, for the observational fields of view $\theta_{\text{FoV}}=1.0^{\circ}$ and $ 4.5^{\circ}$. We find that for the plasma-instability parameters $\lambda_0 = 120~ \text{kpc}$, $\alpha = -0.5$, and $\tilde{E} = 1.0~\text{TeV}$, the minimum magnetic field strength required to remain consistent with the Fermi-LAT data is $B\gtrsim2.7\times10^{-17}~\text{G}$, within $\theta_{\text{FoV}}=1.0^{\circ}$.

\section{Constraint on IGMF from other blazar sources}\label{sec:energy-spectrum-other-sources}
\begin{figure*}[!ht]
   \centering
   \includegraphics[height=0.33\textwidth]{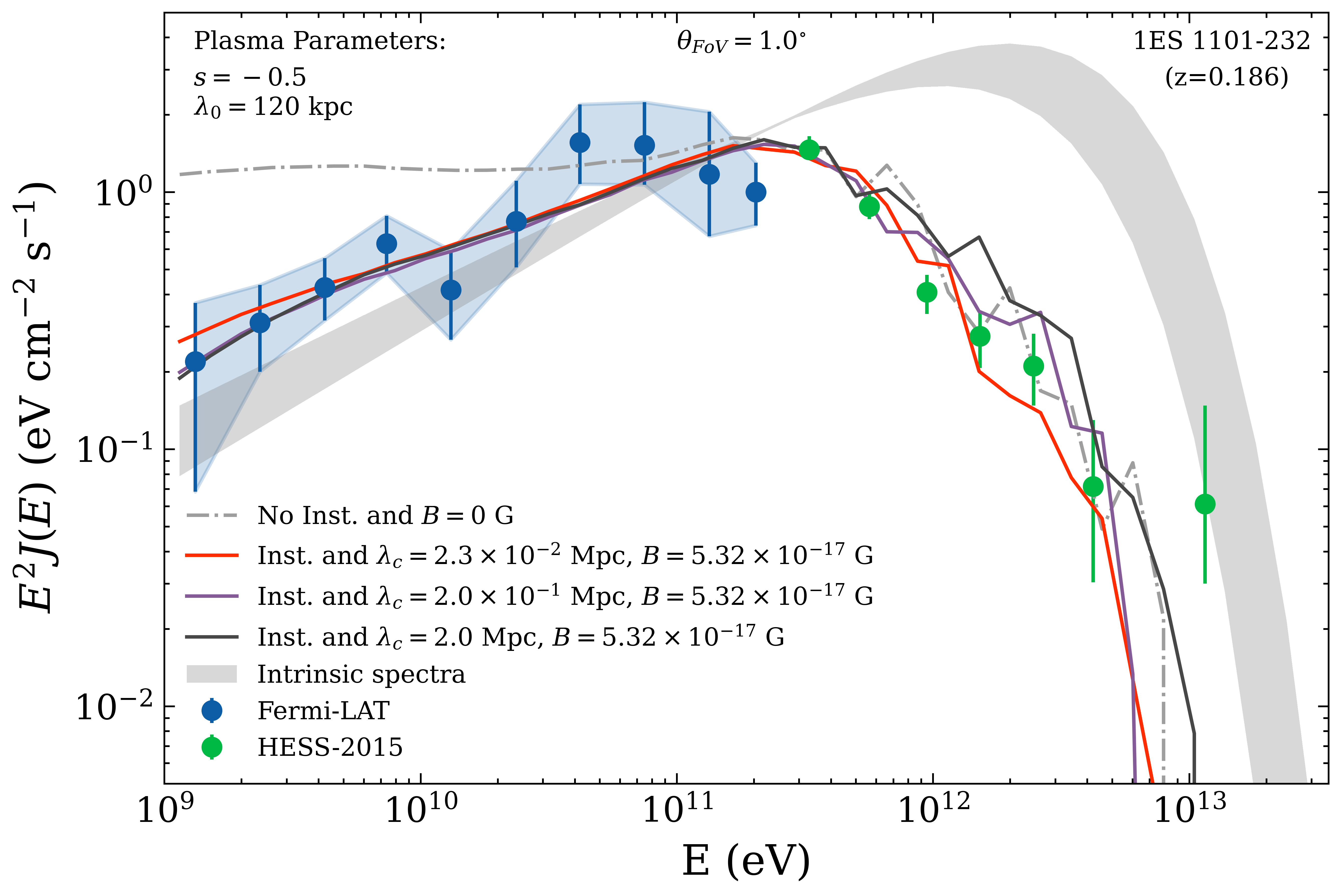}
   \includegraphics[height=0.33\textwidth]{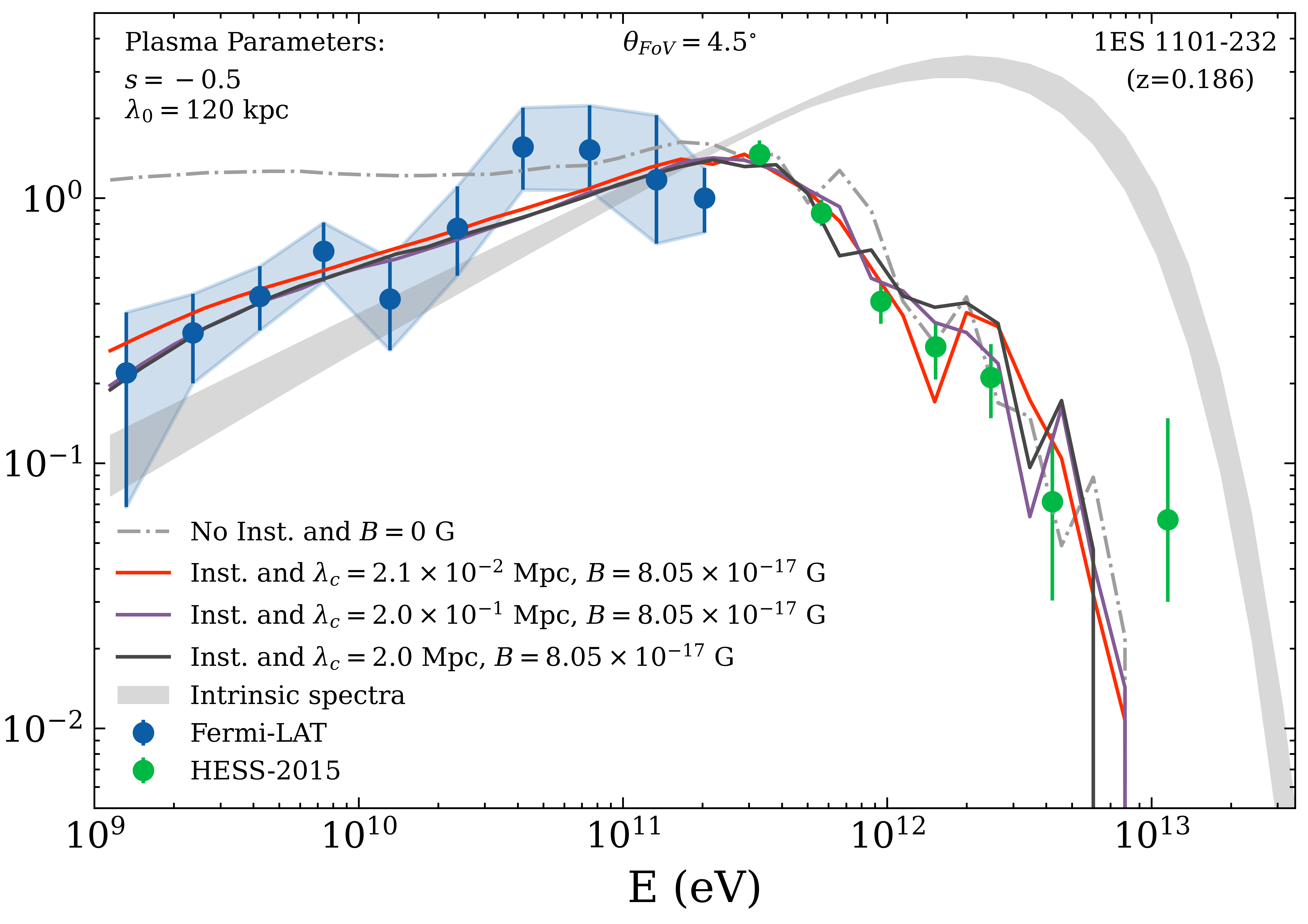}
   \includegraphics[height=0.33\textwidth]{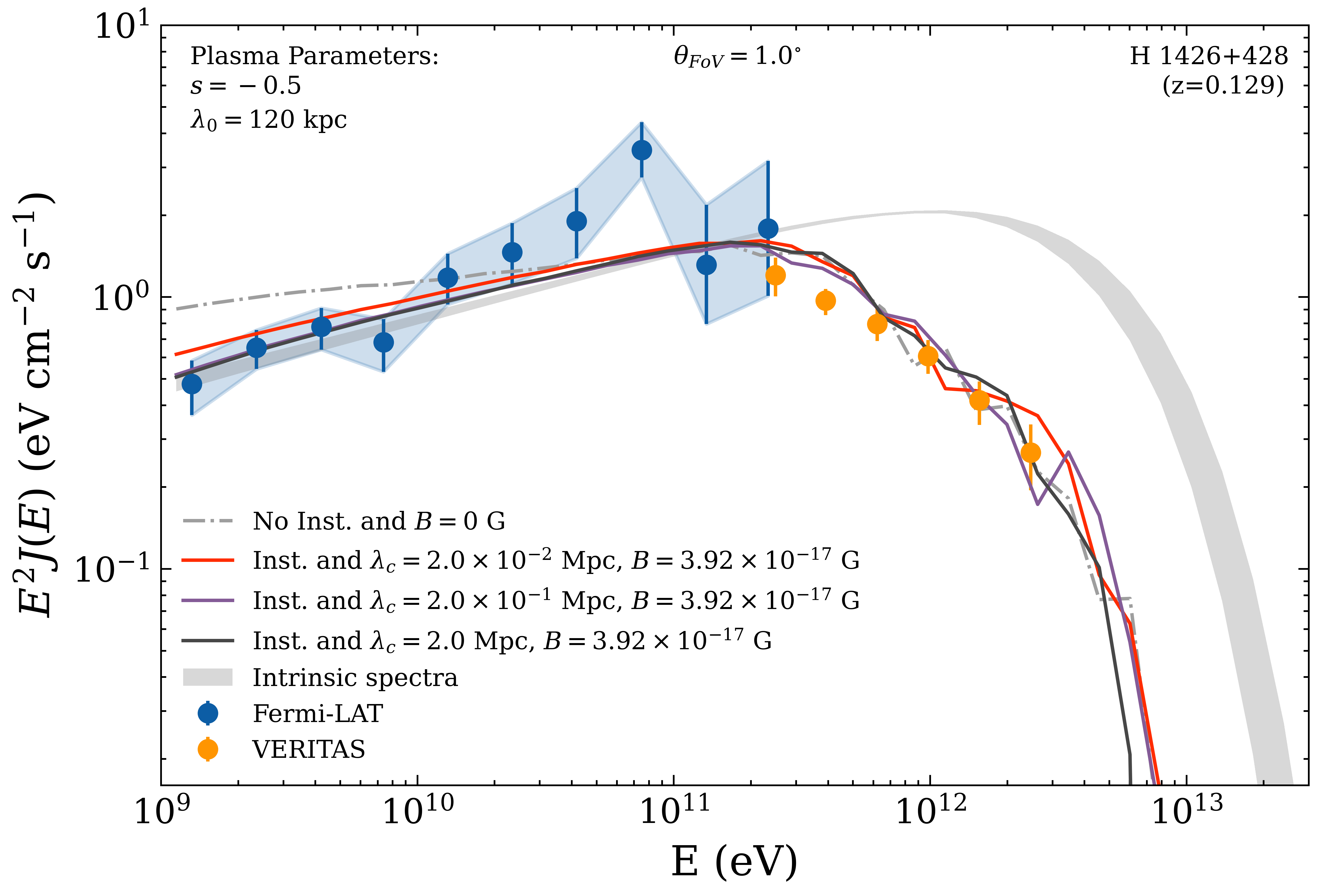}
   \includegraphics[height=0.33\textwidth]{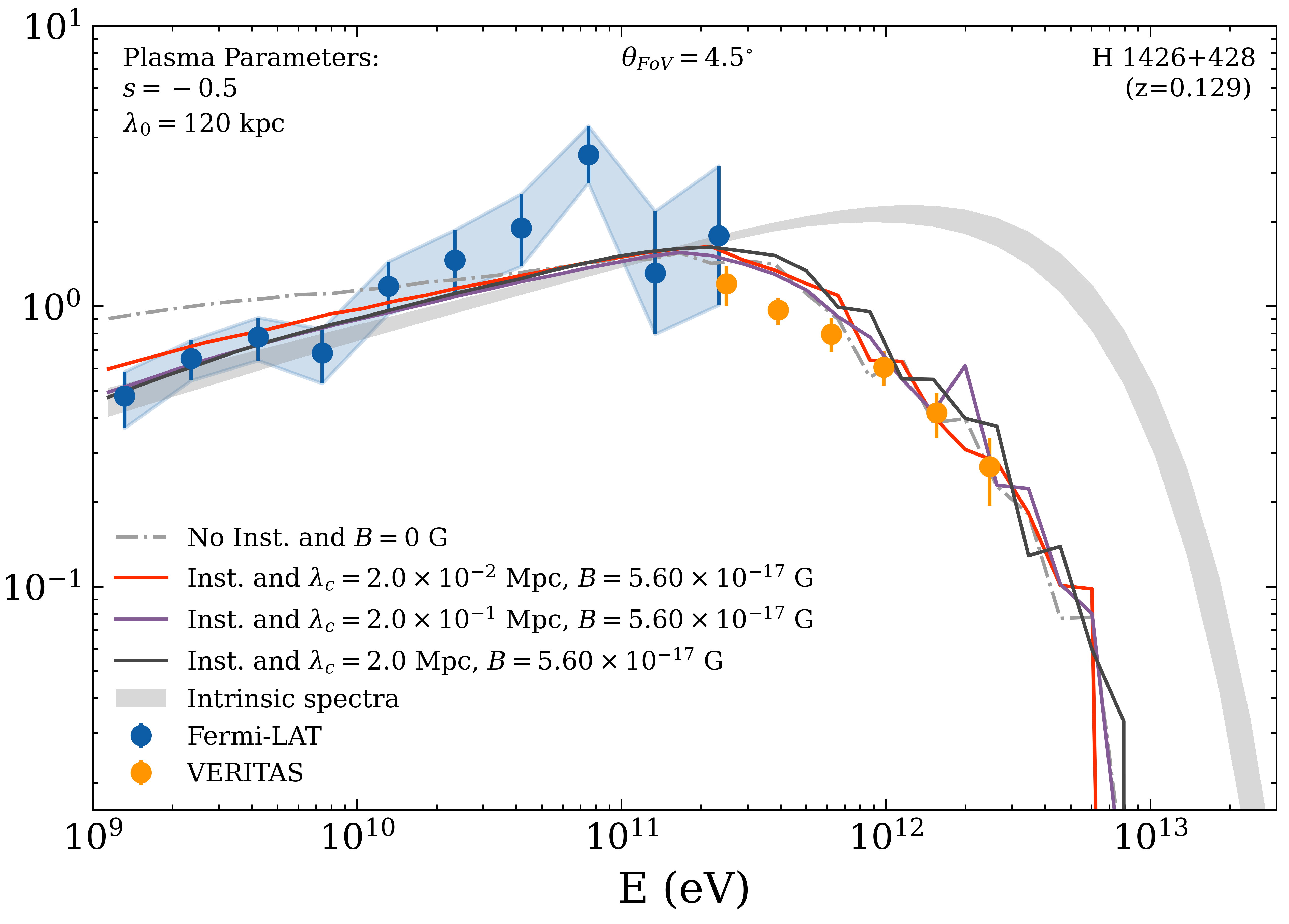}
   \includegraphics[height=0.33\textwidth]{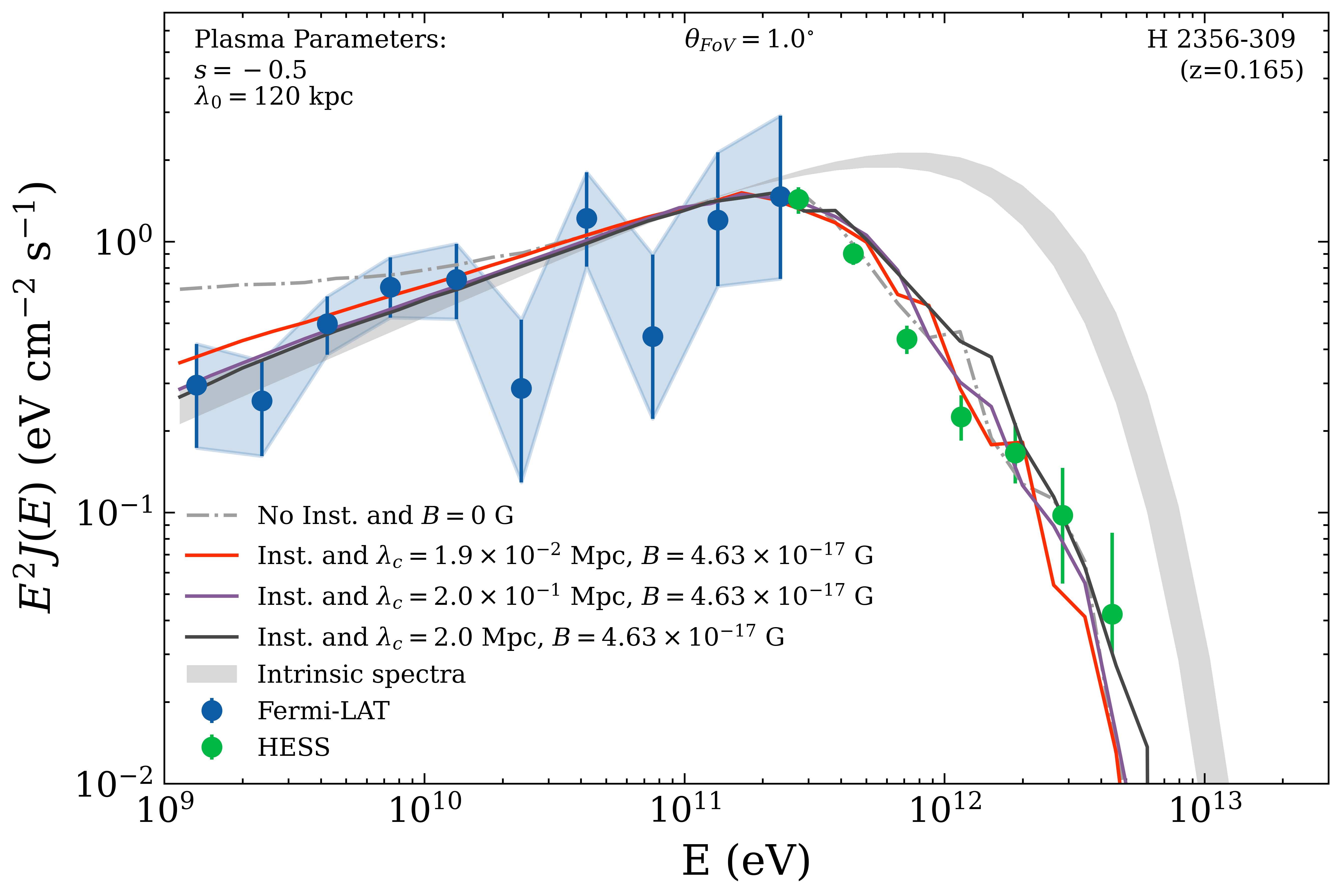}
   \includegraphics[height=0.33\textwidth]{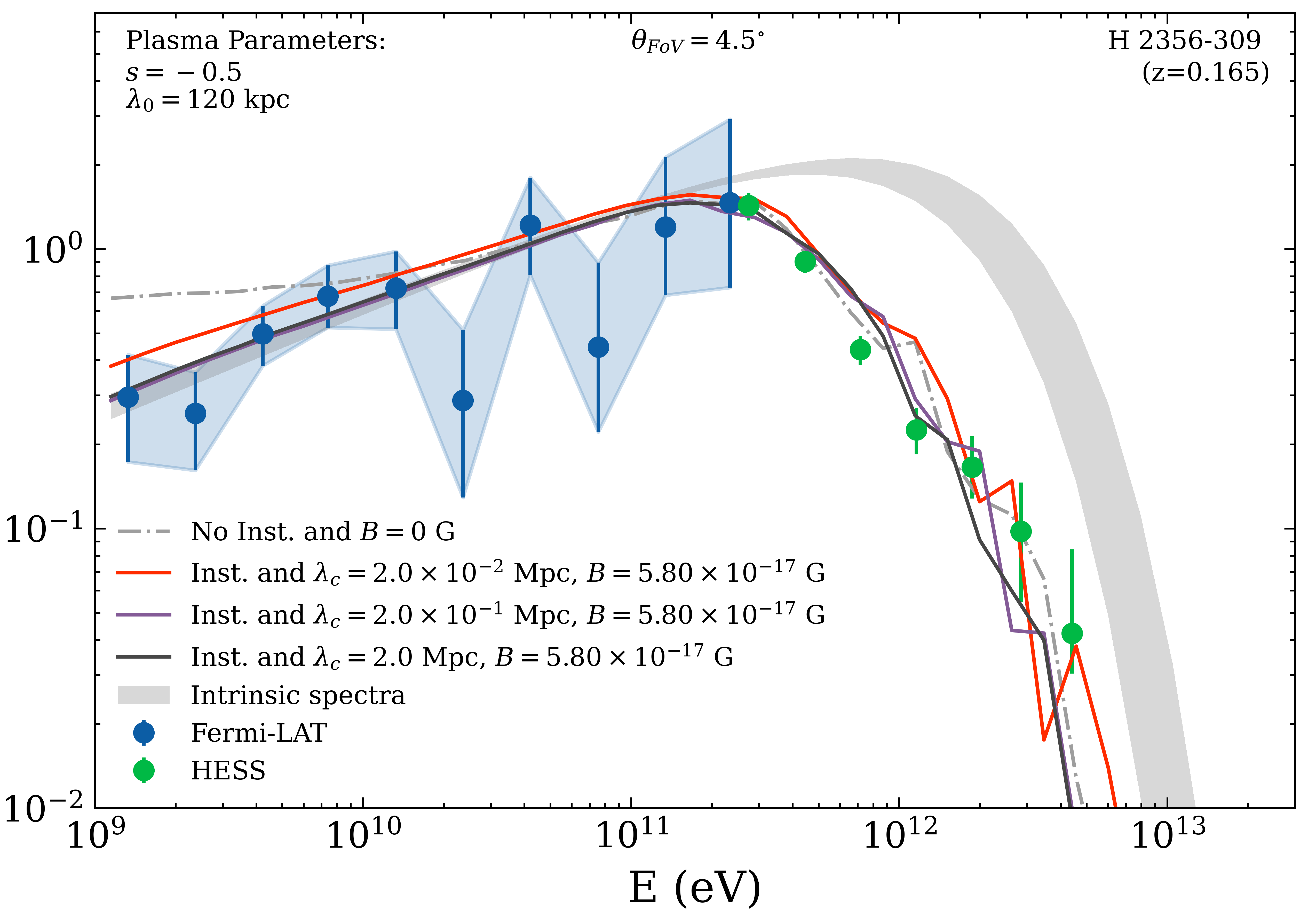}
      \caption{Energy spectra of 1ES 1101–232 ($z \sim 0.186$; top row), H 1426+428 ($z \sim 0.129$; middle row), and H 2356–309 ($z \sim 0.165$; bottom row) in the energy range $10^{-3}\leq E/\text{TeV} \leq 10^{2}$. The gray dash-dotted curve shows the propagated photon spectrum without plasma-instability cooling and without IGMF. Colored solid curves represent propagated spectra including plasma instability cooling with $\lambda_{0} = 120\text{~kpc}$ and $\alpha = -0.5$, for different combinations of IGMF strength and coherence length. The gray band shows the intrinsic spectra, while the corresponding injection parameters are listed in Table~\ref{tab:chi-sq2-other-sources}. Blue data points correspond to Fermi-LAT observations in all panels \cite{HESS:2023zwb, Fermi-LAT:2009pff}. Green data points show H.E.S.S. observations for 1ES 1101--232 and H 2356--309 \cite{HESS:2023zwb}, while orange data points show VERITAS observations for H 1426+428 \cite{farrell2022veritas}. The left and right columns correspond to fields of view $\theta_{\text{FoV}} = 1.0^{\circ}$ and $4.5^{\circ}$, respectively.}
              \label{fig:other-sources}%
\end{figure*}

We extend our analysis to three additional blazar sources for which the data quality is sufficient to detect secondary emission and to derive constraints on IGMF. The blazar 1ES 1101--232 ($z\sim 0.186$) is used by \cite{Neronov:2010gir} in an early analysis to derive constraints on IGMF in the void. H 1426+428 ($z\sim 0.129$) is examined by \cite{Blunier:2025ddu} to derive a lower bound on the IGMF strength, while H 2356--309 ($z\sim 0.165$) is considered in the IGMF analysis by \cite{HESS:2023zwb}. We explore the parameter space over $B \in [1.0,10.0]\times10^{-17}\text{~G}$ with a step size 
$\Delta B = 0.01\times10^{-17}\text{~G}$, and 
$\lambda_{\text{c}} \in [10^{-2},10^{1}]\text{~Mpc}$ with a step size $\Delta \log_{10}\lambda_{\text{c}} = 3 \times 10^{-3}\text{~Mpc}$ for both cases $\theta_{\text{FoV}} = 1.0^\circ$ and $4.5^\circ$. Fig.~\ref{fig:other-sources} shows the energy spectra of 1ES 1101--232 in the top row, H 1426+428 in the middle row, and H 2356--309 in the bottom row. The left and right columns correspond to fields of view $\theta_{\text{FoV}} = 1.0^{\circ}$ and $4.5^{\circ}$, respectively.

\begin{table*}[!ht]\footnotesize\centering
\caption{$\chi^{2}_\text{min}(A, \beta, E_{\text{cut}};\lambda_{\text{c}}, B)/n_{\text{dof}}$ values are obtained for the Fermi-LAT ($n_{\text{dof}}=7$) data, $\theta_{\text{FoV}}=1.0^{\circ}$ and $4.5^{\circ}$ and best-fit plasma instability parameters, $\lambda_{0}=120$ kpc, $\alpha=-0.5$. The $\chi^2_{\min}/n_{\text{dof}}
= \min_{\lambda_{\text{c}}, B} [ \chi^2_{\min}(A, \beta, E_{\text{cut}};\lambda_{\text{c}}, B) ]/n_{\text{dof}}$ value is highlighted in bold.}\label{tab:chi-sq2-other-sources}
\resizebox{1.0\textwidth}{!}{
\begin{tabular}{c*{9}{c}r}
 \hline
 \multirow{2}{*}{$\lambda_{0}$ [kpc]} & \multirow{2}{*}{$\alpha$} & \multirow{2}{*}{Source} & \multirow{2}{*}{$\theta_{\text{FoV}}$ [deg]} & \multirow{2}{*}{$\text{B}/10^{-17}~\text{[G]}$} & \multirow{2}{*}{$\lambda_{\text{c}}$ [Mpc]} & \multirow{2}{*}{$\beta$} & \multirow{2}{*}{$E_{\text{cut}}$ [TeV]} & $\text{log}_{10}A$ & \multirow{2}{*}{$\chi_{\text{min}}^{2}/n_{\text{dof}}$}\\ & & & & & & & & \makecell[c]{[eV$^{-1}$cm$^{-2}$sec$^{-1}$]} \\ \hline
{} & {} & {} & {} & {} & $2.3\times 10^{-2}$ & $1.51$ & $2.2$ & $-23.390$ & $\textcolor{black}{1.175}$ \\
{} & {} & {} & $1.0$ & $5.32$ & $\mathbf{2.0\times 10^{-1}}$ & $\mathbf{1.45}$ & $\mathbf{2.8}$ & $\mathbf{-23.371}$ & $\textcolor{black}{\mathbf{1.129}}$ \\
{} & {} & \multirow{2}{*}{1ES 1101--232} & {} & {} & $\mathbf{2.0}$ & $\mathbf{1.40}$ & $\mathbf{3.2}$ & $\mathbf{-23.340}$ & $\textcolor{black}{\mathbf{1.129}}$ \\ \cline{4-10}
{} & {} & {} & {} & {} & $2.1\times 10^{-2}$ & $1.51$ & $3.6$ & $-23.453$ & $\textcolor{black}{0.967}$ \\
{} & {} & {} & $4.5$ & $8.05$ & $\mathbf{2.0\times 10^{-1}}$ & $\mathbf{1.41}$ & $\mathbf{3.5}$ & $\mathbf{-23.391}$ & $\textcolor{black}{\mathbf{0.844}}$ \\
{} & {} & {} & {} & {} & $\mathbf{2.0}$ & $\mathbf{1.43}$ & $\mathbf{4.1}$ & $\mathbf{-23.434}$ & $\textcolor{black}{\mathbf{0.844}}$ \\ \cline{3-10}
{} & {} & {} & {} & {} & $2.0\times 10^{-2}$ & $1.76$ & $4.5$ & $-23.585$ & $\textcolor{black}{2.349}$ \\
{} & {} & {} & $1.0$ & $3.92$ & $\mathbf{2.0\times 10^{-1}}$ & $\mathbf{1.74}$ & $\mathbf{3.9}$ & $\mathbf{-23.580}$ & $\textcolor{black}{\mathbf{1.881}}$ \\
\multirow{2}{*}{$120$} & \multirow{2}{*}{$-0.5$} & \multirow{2}{*}{H 1426+428} & {} & {} & $\mathbf{2.0}$ & $\mathbf{1.73}$ & $\mathbf{3.2}$ & $\mathbf{-23.553}$ & $\textcolor{black}{\mathbf{1.881}}$ \\ \cline{4-10}
{} & {} & {} & {} & {} & $2.0\times 10^{-2}$ & $1.76$ & $3.8$ & $-23.586$ & $\textcolor{black}{2.333}$ \\
{} & {} & {} & $4.5$ & $5.60$ & $\mathbf{2.0\times 10^{-1}}$ & $\mathbf{1.73}$ & $\mathbf{4.6}$ & $\mathbf{-23.583}$ & $\textcolor{black}{\mathbf{1.840}}$ \\
{} & {} & {} & {} & {} & $\mathbf{2.0}$ & $\mathbf{1.71}$ & $\mathbf{4.3}$ & $\mathbf{-23.541}$ & $\textcolor{black}{\mathbf{1.840}}$ \\ \cline{3-10}
{} & {} & {} & {} & {} & $1.9\times 10^{-2}$ & $1.64$ & $1.6$ & $-23.483$ & $\textcolor{black}{2.058}$ \\
{} & {} & {} & $1.0$ & $4.63$ & $\mathbf{2.0\times 10^{-1}}$ & $\mathbf{1.60}$ & $\mathbf{1.4}$ & $\mathbf{-23.442}$ & $\textcolor{black}{\mathbf{1.551}}$ \\
{} & {} & \multirow{2}{*}{H 2356--309} & {} & {} & $\mathbf{2.0}$ & $\mathbf{1.58}$ & $\mathbf{1.8}$ & $\mathbf{-23.438}$ & $\textcolor{black}{\mathbf{1.551}}$ \\ \cline{4-10}
{} & {} & {} & {} & {} & $2.0\times 10^{-2}$ & $1.64$ & $1.9$ & $-23.459$ & $\textcolor{black}{2.677}$ \\
{} & {} & {} & $4.5$ & $5.80$ & $\mathbf{2.0\times 10^{-1}}$ & $\mathbf{1.60}$ & $\mathbf{1.3}$ & $\mathbf{-23.432}$ & $\textcolor{black}{\mathbf{1.615}}$ \\
{} & {} & {} & {} & {} & $\mathbf{2.0}$ & $\mathbf{1.61}$ & $\mathbf{1.2}$ & $\mathbf{-23.435}$ & $\textcolor{black}{\mathbf{1.615}}$ \\ \hline
\end{tabular}}
\end{table*}
\begin{table*}[!ht]
\centering
\caption{Summary of constraints on the IGMF for other blazar sources.}
\label{tab:igmf-limit-other-sources}
\resizebox{0.55\textwidth}{!}{
\begin{tabular}{c c c c c}
\hline
\multirow{2}{*}{Sources} & \multirow{2}{*}{$z$} & \multirow{2}{*}{$\lambda_{\text{c}}\text{~[Mpc]}$} & \multicolumn{2}{c}{$\text{B}/10^{-17}~\text{[G]}$} \\ 
\cline{4-5}
 & & & \makecell[c]{$\theta_{\text{FoV}} = 1.0^{\circ}$} & \makecell[c]{$\theta_{\text{FoV}} = 4.5^{\circ}$} \\ 
\hline
1ES 1101--232 & 0.186 & {} & 5.32 & 8.05 \\
H 1426+428   & 0.129 & 0.2 & 3.92 & 5.60 \\
H 2356--309  & 0.165 & {} & 4.63 & 5.80 \\
\hline
\end{tabular}}
\end{table*}

We fit the total spectra of 1ES 1101--232 and H 2356--309 using the Fermi-LAT and H.E.S.S. data sets used in \cite{HESS:2023zwb}, and that of H 1426+428 using the Fermi-LAT and VERITAS data reported by \cite{Fermi-LAT:2009pff, farrell2022veritas}. Here we perform the same $\chi^2$ analysis as used for the source 1ES 0229+200 to obtain the best-fit scenario, as shown in Table \ref{tab:chi-sq2-other-sources}. \textcolor{black}{The minimum $\chi^2/n_{\text{dof}}$ values in Table \ref{tab:chi-sq2-other-sources} are generally close to unity, suggesting that the explored parameter combinations provide a good agreement between the predicted spectra and the Fermi-LAT observational data.} Table \ref{tab:igmf-limit-other-sources} summarizes the constraints on the IGMF derived from the three sources considered. The constraints on the IGMF from additional sources are broadly consistent with those obtained from the primary analysis of 1ES 0229+200. In particular, the derived IGMF limits remain in the range of $[3.9-8.1]\times 10^{-17}~\text{G}$, with a similar dependence on the fields of view and plasma instability cooling.

\section{Discussion and conclusion}\label{sec:disscussion}
In this study, we conduct a parametric analysis of energy losses for pair beams driven by plasma instabilities in the presence of IGMF. We perform an extragalactic propagation of high-energy photons from the blazar 1ES 0229+200 using a Monte Carlo simulation framework \texttt{CRPropa 3.2}. We obtain the propagated photon spectrum for both the IGMF-influenced cascade and the cascade modified by plasma instabilities, using the size of the extended emission as our analysis method. We construct the cascade signal matrix as a function of the plasma parameters, the IGMF strength and coherence length, and the observer’s field-of-view angle. We then minimize the $\chi^{2}/n_\text{dof}$ for both scenarios to constrain the IGMF that best reproduces the observed photon spectrum with a $2\sigma$ confidence level. We evaluate the test statistic using only the GeV-band Fermi-LAT data, since the cascade emission contributes most significantly at these energies.

We find that, for the instability parameters $\lambda_{0} = 120~\text{kpc}$ and $\alpha = -0.5$ consistency with the Fermi-LAT observations requires a magnetic-field strength of at least $B \gtrsim 2.7 \times 10^{-17}~\text{G}$ for an observer field of view $\theta_{\text{FoV}} = 1.0^\circ$.
When the instability cooling lengths is an order of magnitude larger than the IC interaction length quench the cascade so rapidly that the production of GeV secondary photons is strongly suppressed even in the absence of an IGMF. In the best-fit scenario, the fractional energy loss through instability within a single IC interaction length is marginal, at $\lesssim 2\%$ for electrons with energies below $4~\text{TeV}$ and $\lesssim 1\%$ for electrons in the GeV range. The lower bound on the IGMF derived by \cite{MAGIC:2022piy} (the red line shown in Fig.~\ref{B-lambda-c}) from combined MAGIC and Fermi-LAT observations using time delay analysis is weaker than the limit we obtain for the blazar 1ES 0229+200 without accounting for plasma instabilities. This difference arises from the distinct characteristic energy scalings of the two methods since the delayed cascade emission follows an energy dependence of $\tau_{\text{delay}}\propto E^{-5/2}$, while the angular extent of the emission varies as $\theta_{\text{obs}}\propto E^{-1}$ for $\lambda_{\text{c}}<l_{\text{IC}}$. The arrival time delay for a $1\text{~GeV}$ photon is approximately 300 times longer than that of a $10\text{~GeV}$ photon, making it more challenging to detect the delayed signal promptly in time delay analyses. Consequently, the $1\text{~GeV}$ photon has an angular spread roughly 10 times larger than that of a $10\text{~GeV}$ photon, making this method more effective for detecting lower-energy photons. 

In Fig.~\ref{B-lambda-c}, we have shown the existing IGMF bounds. Cosmological magnetic fields likely originated during the electroweak and quantum chromodynamics (QCD) phase transitions or inflation \cite{Grasso:2000wj, Durrer:2013pga}. Their initial coherence length is limited by the cosmological horizon size at generation, approximately 100 au for electroweak and 1 parsec for QCD fields. Turbulent decay from generation to recombination reduces field strength but increases coherence length up to the scale of the largest eddies $\sim (B/10^{-8}\text{~G})\text{~Mpc}$ \cite{Banerjee:2004df}. Magnetic reconnection may lead to somewhat shorter lengths. GRB pair-echo data constrain the field to $B \gtrsim 10^{-15}~\text{G,~}\lambda_{\text{c}}\gtrsim 0.1\text{~pc}$ (indicated as the purple dashed line in Fig.~\ref{B-lambda-c}) \cite{Banerjee:2004df}. Ultra-high-energy cosmic ray (UHECR) observations provide the upper bounds on IGMF. This limit on the IGMF strength is around $B\sim 10^{-10}\text{~G}$ for fields with coherence lengths exceeding the size of a supercluster, approximately $70\text{~Mpc}$ \cite{Neronov:2021xua}, as indicated by the light brown shaded region in Fig.~\ref{B-lambda-c}. The lower bound on the IGMF derived from LHAASO and Fermi-LAT observations of GRB 221009A (shown as the black line in the same figure), derived by \cite{Vovk:2023qfk}, is approximately $B\sim 10^{-19}\text{~G}$. In a subsequent study, \cite{Burmeister:2025lgo} derived an improved constraint from the Fermi-LAT and LHAASO observations of the source GRB 221009A, excluding $B < 2.5 \times 10^{-17}~\text{G}$ at 95\% confidence level for $\lambda_{\text{c}}\gtrsim 1~\text{Mpc}$. These results are obtained using a time delay analysis method similar to that employed by \cite{MAGIC:2022piy}. The difference between this limit and existing bounds arises from the distinct energy ranges in which the time-delayed cascade emission was searched for these two sources.

In addition to the IGMF constraint derived from 1ES 0229+200, we extend the analysis to three additional blazars, 1ES 1101--232, H 1426+428, and H 2356--309. The available data for these sources are sufficient to detect the cascade flux. The IGMF limits derived from these sources are consistent with those obtained from 1ES 0229+200. In general, the field strengths are at the level of a few $\times 10^{-17}~\text{G}$, with a similar dependence on the fields of view and plasma instability cooling. We find that plasma-instability cooling alone can reproduce the observed spectra within the statistical uncertainties; the addition of a nonzero weak IGMF yields a slightly better improvement in the normalized $\chi^2$ values than the instability-only scenario.
 
Future studies will address the limitation of assuming spatially uniform energy losses, in which all electrons lose energy regardless of their distance from the source. Introducing a distance-dependent instability cooling could improve the understanding of the cooling dynamics. Additionally, a recent study \cite{Alawashra:2025tcj} suggests that MeV cosmic-ray electrons can suppress the TeV pair-beam plasma instability via linear Landau damping. Exploring this effect within our framework could offer valuable insights into the instability behavior.
\section*{Data availability}
All external data used in this study were obtained from the cited sources. The data produced and analyzed during this work are available from the corresponding author upon reasonable request.

\section*{CRediT authorship contribution statement}
\textbf{Suman Dey:} Writing – original draft, Writing – review \& editing, Software, Methodology, Investigation, Formal analysis, Data curation, Conceptualization, Visualization, Validation.
\textbf{Simone Rossoni:} Writing – review \& editing, Methodology, Investigation, Visualization, Validation.
\textbf{G\"{u}nter Sigl:} Writing – review \& editing, Methodology, Conceptualization, Supervision, Resources, Visualization, Validation.

\section*{Declaration of competing interest}
The authors confirm that they have no competing financial interests or personal relationships that could have influenced the work presented in this paper.

\section*{Acknowledgments}
SD was funded by the Deutsche Forschungsgemeinschaft (DFG, German Research Foundation) under Germany’s Excellence Strategy– EXC 2121 “Quantum Universe”– 390833306. This project was conceived by GS. The authors gratefully acknowledge the use of the CRPropa 3.2 framework for the numerical simulations performed in this work. The authors acknowledge the HPC facility of the Phenod cluster operated at Deutsches Elektronen-Synchrotron (DESY), Hamburg, Germany, and the PHYSnet computational resources operated at II. Institut f\"{u}r Theoretische Physik, Universit\"{a}t Hamburg. We are grateful to the reviewer for the constructive input.

\appendix

\section{Electromagnetic cascade with plasma instability only}\label{appen:cascade-with-instability-only}
In order to reproduce the observed photon spectrum of the blazar 1ES 0229+200, we now investigate the cascade emission affected by plasma instabilities only.
\begin{figure*}[!ht]
   \centering
   \includegraphics[width=0.49\textwidth]{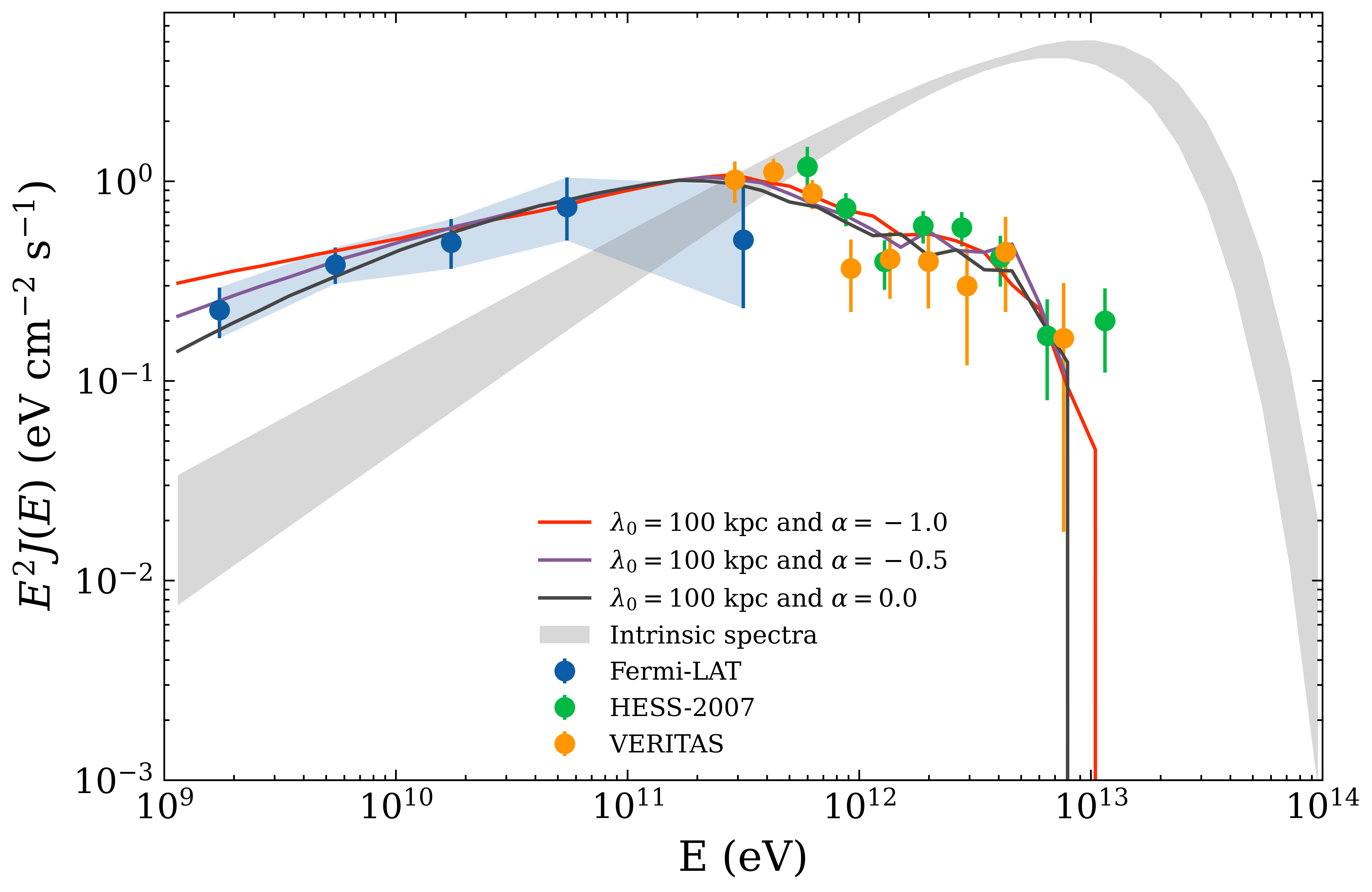}
   \includegraphics[width=0.49\textwidth]{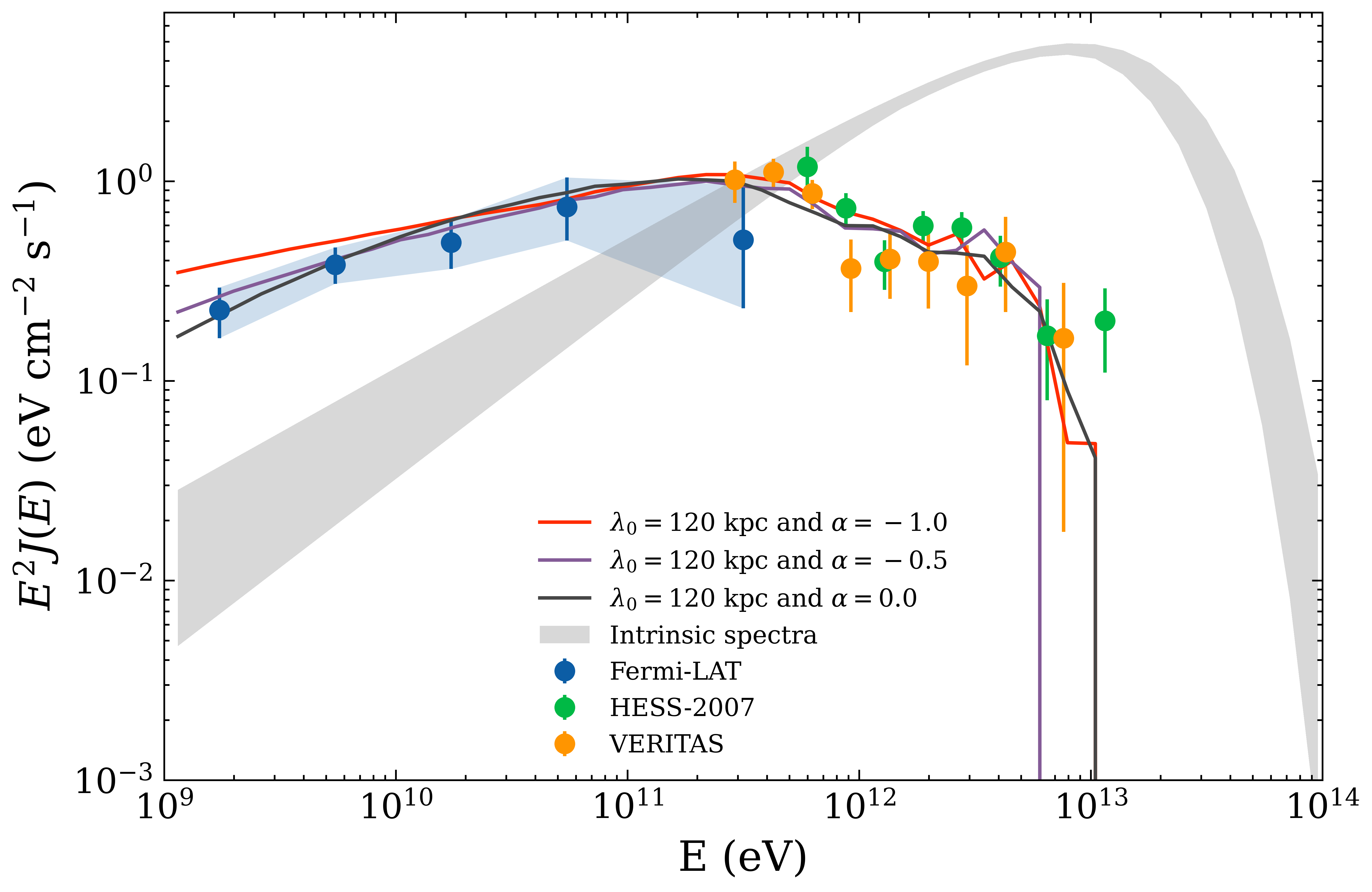}
   \includegraphics[width=0.49\textwidth]{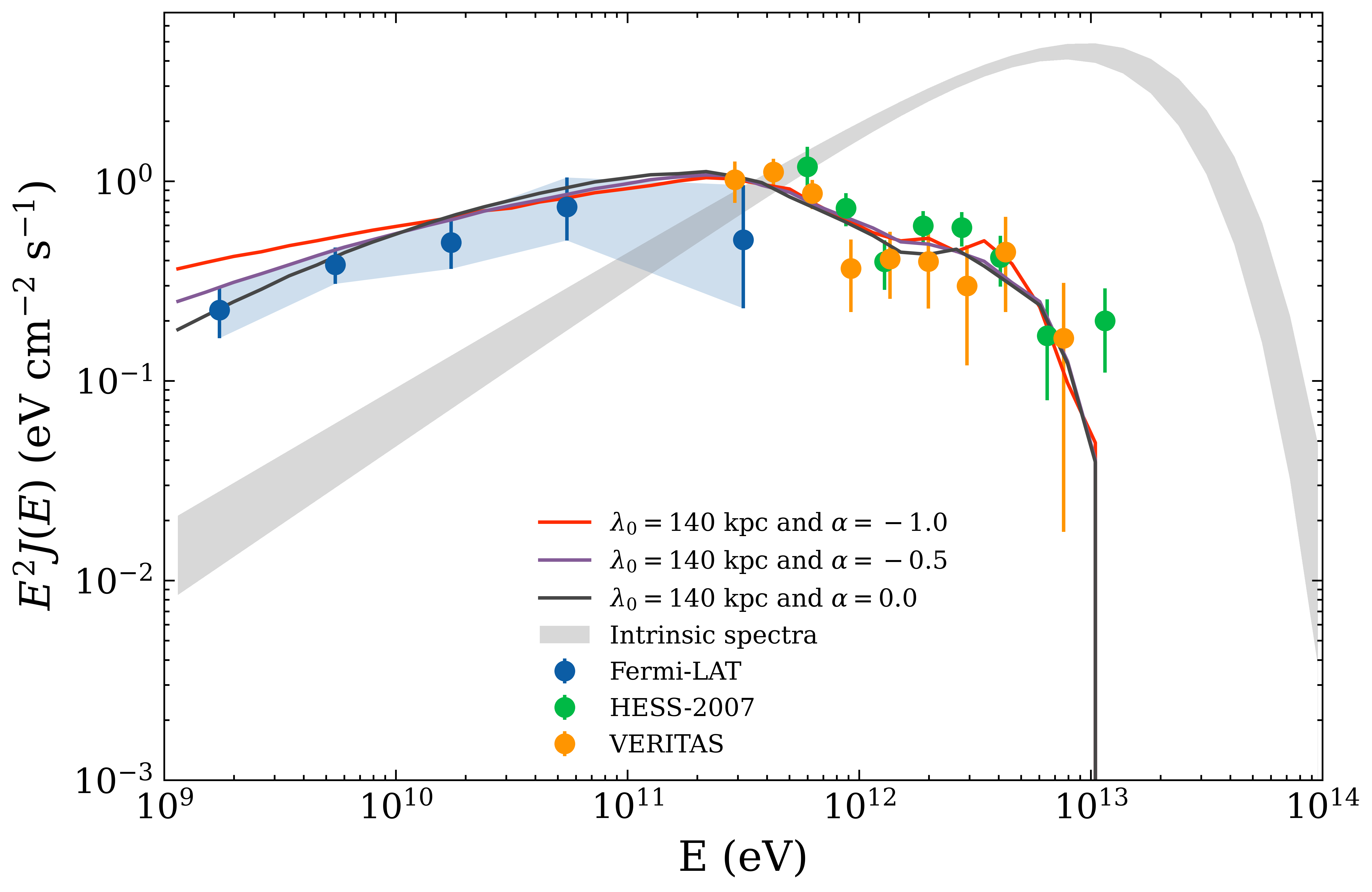}
   \includegraphics[width=0.49\textwidth]{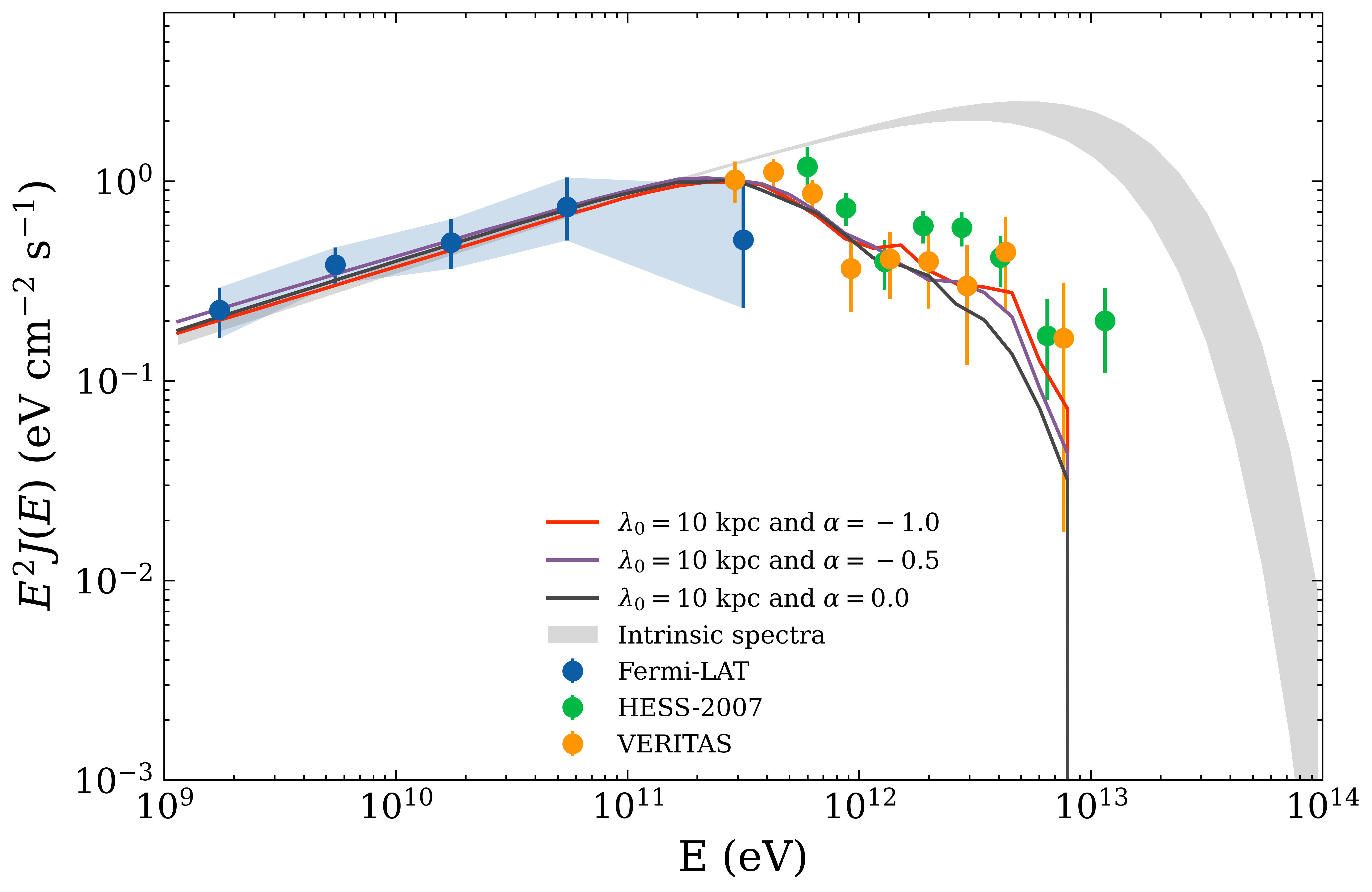}
   \caption{The energy spectrum of 1ES 0229+200 in the energy range $10^{-3}\leq E/\text{TeV} \leq 10^{2}$. The colored solid curves represent the propagated photon spectra for different combinations of the instability parameters, $\lambda_{0}$ and $\alpha$, in the absence of an IGMF, following an approach similar to that of \cite{Castro:2024ooo}. The blue data points indicate the Fermi-LAT \cite{MAGIC:2022piy}, the green data points show the H.E.S.S. \cite{HESS:2007xak}, and the orange data points show the VERITAS \cite{Aliu:2013pya} spectrum. The gray band represents the intrinsic spectra, and the corresponding injection parameters are listed in Table~\ref{tab:chi-sq2-only-inst}. We investigate the energy spectrum considering only the instability-cooling effect, which is consistent with the Fermi-LAT, H.E.S.S., and VERITAS observational data in each plot. The $\chi^{2}/n_{\text{dof}}$ test statistic obtained for the GeV-band data from Fermi-LAT ($n_{\text{dof}}=2$) is also listed in Table~\ref{tab:chi-sq2-only-inst}.}
              \label{fig:spectra-with-only-instability-parameters}%
\end{figure*}
\begin{table*}[!ht]
\footnotesize
\centering
\caption{$\chi^{2}_\text{min}(A, \beta, E_{\text{cut}};\lambda_{0}, \alpha)/n_{\text{dof}}$ values are determined for the GeV-band data from Fermi-LAT ($n_{\text{dof}}=2$), for a combination of instability parameters, $\lambda_{0}$ and $\alpha$. The minimum $\chi^{2}_{\text{min}}/n_{\text{dof}} = \min_{\lambda_{0}, \alpha} [ \chi^2_{\min}(A, \beta, E_{\text{cut}};\lambda_{0}, \alpha) ]/n_{\text{dof}}$ values are highlighted in bold.}
\label{tab:chi-sq2-only-inst}
\resizebox{0.60\textwidth}{!}{
\begin{tabular}{c c c c c c}
 \hline
 \multirow{2}{*}{$\lambda_{0}$ [kpc]}
 & \multirow{2}{*}{$\alpha$}
 & \multirow{2}{*}{$\beta$} 
 & \multirow{2}{*}{$E_{\text{cut}}$ [TeV]} 
 & $\log_{10}A$ 
 & \multirow{2}{*}{$\chi_{\text{min}}^{2}/n_{\text{dof}}$} \\
 & & & &\makecell[c]{[eV$^{-1}$cm$^{-2}$sec$^{-1}$]} & \\ 
 \hline

{} & $-1.0$ & $1.37$ & $12.4$ & $-23.621$ & $2.242$ \\
$100$ & $\mathbf{-0.5}$ & $\mathbf{1.21}$ & $\mathbf{11.7}$ & $\mathbf{-23.713}$ & $\mathbf{2.049}$ \\
{} & $0.0$ & $1.18$ & $8.4$ & $-23.712$ & $2.119$ \\ \cline{0-5}

{} & $-1.0$ & $1.35$ & $13.3$ & $-23.636$ & $2.467$ \\
$\mathbf{120}$ & $\mathbf{-0.5}$ & $\mathbf{1.26}$ & $\mathbf{10.7}$ & $\mathbf{-23.710}$ & $\mathbf{1.699}$ \\
{} & $0.0$ & $1.11$ & $7.8$ & $-23.711$ & $2.543$ \\ \cline{0-5}

{} & $-1.0$ & $1.32$ & $13.9$ & $-23.677$ & $2.571$ \\
$140$ & $\mathbf{-0.5}$ & $\mathbf{1.29}$ & $\mathbf{10.9}$ & $\mathbf{-23.702}$ & $\mathbf{2.023}$ \\
{} & $0.0$ & $1.21$ & $9.8$ & $-23.749$ & $3.021$ \\ \cline{0-5}

{} & $-1.0$ & $1.62$ & $13.4$ & $-23.703$ & $2.027$ \\
$10$ & $\mathbf{-0.5}$ & $\mathbf{1.65}$ & $\mathbf{12.7}$ & $\mathbf{-23.708}$ & $\mathbf{2.005}$ \\
{} & $0.0$ & $1.64$ & $8.4$ & $-23.711$ & $2.010$ \\

\hline
\end{tabular}}
\end{table*}
Fig.~\ref{fig:spectra-with-only-instability-parameters} shows the energy spectra obtained by varying only the plasma instability parameters, $\lambda_{0}$ and $\alpha$, in the absence of an IGMF. As discussed in Section~\ref{sec:plasma-inst}, the oblique instability dominates in warm plasma beams, for which the typical values of the instability spectral index fall within the negative range, $\alpha \in [-1.0,0.0]$. We therefore explore the former parameter space using a step size of $\Delta\alpha=-0.5$, and $\lambda_{0}\in[100,140]~\text{kpc}$ with a step size of $\Delta\lambda_{0}=20~\text{kpc}$. The best-fit intrinsic parameter scan is performed at the 90\% confidence level using the method described in Subsection~\ref{subsec:test-stat}. The standard $1.64\sigma$ is defined by $\Delta\chi^{2}=\chi^{2}-\chi^{2}_{\text{min}}\leq 4.61$ when two parameters are varied. The corresponding normalized condition is $\Delta\chi^{2}/n_{\text{dof}}\leq 2.3$, which represents the standard normalized threshold for a $1.64\sigma$ deviation with $n_{\text{dof}}=2$. The minimum $\chi^{2}_{\text{min}}(A, \beta, E_{\text{cut}};\lambda_{0},\alpha)/n_{\text{dof}}$ for each propagated spectrum corresponding to a combination of instability parameters, $\lambda_{0}$ and $\alpha$ are listed in Table \ref{tab:chi-sq2-only-inst}. The instability parameters that best match the Fermi-LAT observations are obtained for $\lambda_{0}=120~\text{kpc}$ and $\alpha=-0.5$ in the absence of an IGMF, consistent with the results of \cite{Castro:2024ooo}. In the next subsection, we will use these best-fit instability parameters to study the modified electromagnetic cascade in the presence of both plasma instabilities and the IGMF. 
Additionally, we have also considered the propagated photon spectrum in the case $\lambda_{0} = 10~\text{kpc}$, as shown in the bottom right panel of Fig.~\ref{fig:spectra-with-only-instability-parameters}. We find that the cascade signal in the GeV energy range is suppressed for these choices of instability parameters. This implies that, when the instability cooling lengths are approximately an order of magnitude larger than the IC interaction length, the electron--positron pairs lose their energy through plasma instabilities so rapidly that they cannot efficiently upscatter CMB photons via inverse Compton interactions. As a result, the electromagnetic cascade is rapidly quenched due to instability cooling, leading to a strong suppression of the secondary GeV photon flux in the absence of an IGMF.

\biboptions{sort&compress}
\bibliographystyle{elsarticle-num} 
\bibliography{ref}

\end{document}